\documentclass[twocolumn,english,aps,pra,reprint, superscriptaddress,showpacs,longbibliography,showkeys]{revtex4-2}
\usepackage{amsmath,amssymb,bbm,mathrsfs,bm,braket,color,graphicx,comment,xcolor, colortbl,dsfont,multirow, adjustbox, booktabs}
\usepackage[colorlinks,citecolor=blue,urlcolor=blue,linkcolor = blue]{hyperref}
\usepackage[mathscr]{euscript}
\usepackage{tikz}
\usetikzlibrary{quantikz2}
\usetikzlibrary{arrows.meta}
\usepackage[normalem]{ulem}

\newcommand{\calC}{\mathcal{C}}
\DeclareMathOperator{\Tr}{Tr}
\newcommand{\bfq}{{\boldsymbol{q}}}
\newcommand{\bfbeta}{{\boldsymbol{\beta}}}
\newcommand{\bfgamma}{\boldsymbol{\gamma}}
\newcommand{\bfOmega}{\boldsymbol{\Omega}}
\newcommand{\eeuler}{\mathrm{e}}
\DeclareMathOperator*{\argmax}{arg\,max}

\newcommand{\rmmin}{\mathrm{min}}
\newcommand{\rmmax}{\mathrm{max}}
\newcommand{\squaredots}{
    \vspace{-.175em}
    \tikz[line cap=round, line join=round]{
    \draw[black] (0ex,0ex) -- (0ex,0.8ex) --  (0.8ex,0.8ex) --  (0.8ex,0ex) -- cycle;
    \draw[color=black, fill=black] (0ex,0ex) circle (0.175ex);
    \draw[color=black, fill=black] (0ex,0.8ex) circle (0.175ex);
    \draw[color=black, fill=black] (0.8ex,0ex) circle (0.175ex);
    \draw[color=black, fill=black] (0.8ex,0.8ex) circle (0.175ex);
    }
}

\newcommand{\parity}{Parity}
\newcommand{\Parity}{Parity}
\newcommand{\vanilla}{vanilla}
\newcommand{\Vanilla}{Vanilla}

\newcommand{\Aclassical}{A_\text{class}^{(p)}}

\begin{document}

\title{Performance of Parity QAOA for the Signed Max-Cut Problem}

\author{Anita Weidinger}
\affiliation{Institut für Theoretische Physik, Universität Innsbruck, A-6020 Innsbruck, Austria}
\author{Glen Bigan Mbeng}
\affiliation{Institut für Theoretische Physik, Universität Innsbruck, A-6020 Innsbruck, Austria}
\affiliation{Parity Quantum Computing GmbH, A-6020 Innsbruck, Austria}
\author{Michael Fellner}
\affiliation{Institut für Theoretische Physik, Universität Innsbruck, A-6020 Innsbruck, Austria}
\affiliation{Parity Quantum Computing GmbH, A-6020 Innsbruck, Austria}
\author{Davit Khachatryan}
\affiliation{Parity Quantum Computing GmbH, A-6020 Innsbruck, Austria}
\author{Wolfgang Lechner}
\affiliation{Institut für Theoretische Physik, Universität Innsbruck, A-6020 Innsbruck, Austria}
\affiliation{Parity Quantum Computing GmbH, A-6020 Innsbruck, Austria}
\affiliation{Parity Quantum Computing Germany GmbH, 20095 Hamburg, Germany} 

\date{\today}

\begin{abstract}
The practical implementation of quantum optimization algorithms on noisy intermediate-scale quantum devices requires accounting for their limited connectivity. As such, the {\parity} architecture was introduced to overcome this limitation by encoding binary optimization problems onto planar quantum chips. 
We investigate the performance of the Quantum Approximate Optimization Algorithm on the {\parity} architecture ({\parity} QAOA) for solving instances of the signed Max-Cut problem on complete and regular graphs. By comparing the algorithms at fixed circuit depth, we demonstrate that {\parity} QAOA outperforms conventional QAOA implementations based on SWAP networks. Our analysis utilizes Clifford circuits to estimate lower performance bounds for {\parity} QAOA for problem sizes that would be otherwise inaccessible on classical computers.
For single layer circuits we additionally benchmark the recursive variant of the two algorithms, showing that their performance is equal. 
\end{abstract}
\maketitle

\section{Introduction}

Given its broad applicability and commercial value, combinatorial optimization is a key area of quantum computation~\cite{Nielsen_Chuang_book2010, Montanaro_Npj2016}. In this domain, the quantum approximate optimization algorithm (QAOA)~\cite{Farhi2014,Blekos_PhysicsReports2024} recently emerged as a strong candidate for demonstrating quantum advantage on noisy intermediate-scale quantum (NISQ) devices~\cite{Preskill_quantum2018,Zhou_Lukin_prx2020, Shaydulin_Pistoia_Science2024}. 

At its core, QAOA is a variational quantum
algorithm (VQA)~\cite{Carezo_Cole_NaturePhys2021} that uses shallow quantum circuits to prepare states, which encode approximate solutions to
optimization problems. However, practical implementations of the algorithm on existing NISQ devices face challenges due to limited hardware connectivity and gate error rates~\cite{Harrigan_NaturePhysics2021,Pelofske_Eidenbenz_NatureComm2023,Pellow-Jarman_Petruccione_ScientificReports2024}. To address these issues, various QAOA variants were introduced~\cite{Blekos_PhysicsReports2024}, aiming at reducing the algorithm's execution times and hardware requirements ~\cite{ Herrman_ScientificReports2022,Glos_npjQuantumInformation2022, Zsolt_IEEE2020, Fuchs_SNComputerScience2021, Bako_Quantum2025}. These range from methods to identify hardware-efficient circuits such as QAOA+~\cite{Hadfield_Algorithms2019}, ADAPT-QAOA~\cite{Zhu_Economou_PRR2022} and CD-QAOA to feedback methods that further optimize the quantum states such as recursive QAOA (RQAOA)~\cite{Bravyi_Tang_PRL2020}.

{\emph{Parity}} QAOA~\cite{Lechner_2015, Lechner2018, Ender2023} is a QAOA variant designed to specifically address qubit connectivity limitations of existing quantum chips. 
On such chips, conventional QAOA implementations, hereinafter referred to as \emph{vanilla} QAOA, involve SWAP-gates and rerouting strategies which are hard to parallelize and drastically increase the circuit depth~\cite{Harrigan_NaturePhysics2021, Hirata_QuantumInfoConf2011_LNN, Mahabubul_Swaroop_2020, Li_Xie_Proceedings2019, Tan_Cong_Proceedings2020, Sivarajah_IOPPublishing2021, Lingling_Proceedings2022, kotil_arXiv2023, crooks_arXiv2018, Weidenfeller_quantum2022, Kivlichan_Babbush_PRL2018}. 
In contrast, {\parity} QAOA distributes the computational load across physical qubits and leverages the chip's planar structure, relying solely on highly parallelizable local gates~\cite{Lechner2018, Unger_Lechner_arXiv2023}. 
In particular, for typical binary optimization problems, parity QAOA requires shallower circuits than vanilla QAOA and, in some instances, also, fewer entangling gates~\cite{Fellner_Quantum2023}. Moreover, the redundant information distributed across multiple physical qubits is useful for mitigating gate errors on NISQ devices~\cite{Weidinger2023}.

Despite its advantages, {\parity} QAOA's planar geometry limits the spread of correlations across the device. This is discussed in Ref.~\cite{Wybo_Leib_Quantum2024}, where the authors observed that, when solving the Sherrington-Kirkpatrick (SK) spin-glass problem, {\parity} QAOA's performance deteriorates as the problem size is increased, whereas {\vanilla} QAOA's performance does not. This known phenomenon, which affects QAOA beyond complete graphs, may result in obstacles to preparing high-quality approximate solutions to optimization problems~\cite {Bravyi_Tang_PRL2020}. 

In this work, we study {\parity} QAOA to quantify better its advantages and disadvantages over  {\vanilla} QAOA. To do so, we compare the algorithm's performances and hardware requirement on the signed Max-Cut problem~\cite{Crowston_Muciaccia_TheoreticalCS2013}, which includes the SK spin-glass considered in Ref.~\cite{Wybo_Leib_Quantum2024} as a special case. We use Clifford pre-optimization to estimate the algorithm's performance and hardware requirement for solving large problem instances, with sizes that go beyond the ones considered in previous studies~\cite{Weidinger2023, Wybo_Leib_Quantum2024}. We find that when a suitable objective function is considered, {\parity} QAOA consistently outperforms {\vanilla} QAOA at fixed circuit depth. Finally, we address the limited spread of correlation in {\parity} QAOA's planar geometry by introducing a {\parity} recursive QAOA (RQAOA). We show that similarly to its vanilla counterpart, {\parity} RQAOA overcomes the obstacles to state preparation introduced by the local geometry.

The remainder of this work is organized as follows. First, we introduce the problems of consideration in Sec.~\ref{sec:signed_maxcut} and review the two different QAOA methods: {\vanilla} QAOA in Sec.~\ref{sec:QAOA} and {\parity} QAOA in Sec.~\ref{sec:ParityQAOA}.  We then present a procedure to pre-optimize QAOA parameters using Clifford circuits in Sec.~\ref{sec:ParityLowerBound}. After that, we present the performance benchmarks of the two QAOA approaches for the signed Max-Cut problem on a complete graph and a 4-regular graph in Sec.~\ref{sec:QAOA_comparison}, where we focus on CNOT depth and CNOT count as metrics.
Sec.~\ref{sec:recursive_qaoa} introduces and benchmarks a {\parity} RQAOA approach. We conclude the manuscript with Sec.~\ref{sec:conclusion} by summarizing our findings and presenting an outlook on potential future research directions in Sec.~\ref{sec:conclusion}.

\begin{figure}
    \centering
    \includegraphics[width=\columnwidth]{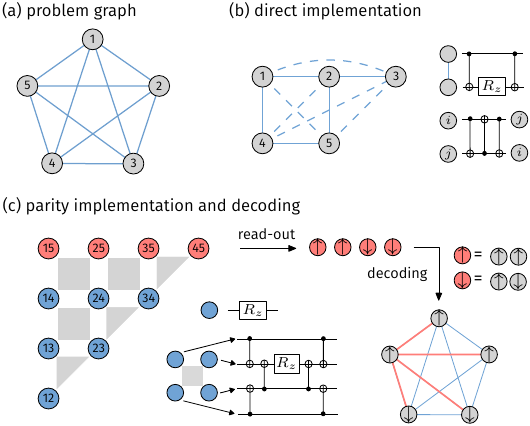}
    \caption{
    (a) Complete graph with $N=5$ nodes. (b) Direct implementation on physical devices with rectangular connectivity. 
    (c) Parity implementation on physical devices with rectangular connectivity. 
     In (b), local interactions are represented by solid blue lines, while non-local interactions, which require SWAP gates to become local, are depicted with dashed blue lines.
    In (c), the interactions are transformed into single qubit $z$-rotations. 
    The constraints are marked as gray squares and triangles. An example of a single 4-body constraint circuit is shown in (c).
    On the right, an example of the decoding procedure is given. Here, $\uparrow$ and $\downarrow$ refer to the measurement outcome $0$ and $1$, respectively. }
    \label{fig:Figure1}
\end{figure}

\section{Background}

\subsection{The Signed Max-Cut problem}\label{sec:signed_maxcut}
This work considers the signed Max-Cut problem, a variation of the classic Max-Cut problem, as introduced in Ref.~\cite{Crowston_Muciaccia_TheoreticalCS2013}, which we define in this section. Possible fields of application for the signed Max-Cut problem are social networks, systems biology, and integrated circuit design~\cite{Chiang_Zelikovsky_IEEE2007,DasGupta_Zhang_Biosystems2007,Huffner_Niedermeier_book2007,Zaslavsky_ElectronicElectronicofCombinatorics2021}. This problem is in the strongly NP-hard complexity class~\cite{Garey_JACM1978,Garey_TCS1976}, therefore, there are no (known) efficient classical optimization algorithms that solve it. In the context of quantum optimization, the signed Max-Cut problem was recently investigated in Ref.~\cite{polloreno_ArXiv2024}.

Let ${G=(V, E)}$ be an undirected graph, with vertices $V$ and edges ${E \subseteq V^2}$. An example of such a graph is depicted in Fig.~\ref{fig:Figure1}. Let each edge ${(i, j) \in E}$ be associated with a weight ${J_{ij} \in \{-1, +1\}}$, indicating whether the connection between vertices $i$ and $j$ is attractive ($+1$) or repulsive ($-1$).
The goal in the signed Max-Cut problem is to find a partition of the vertex set $V$ into two disjoint subsets, $V_1$ and $V_2$, such that ${V_1 \cup V_2 = V}$, with the objective to maximize the sum of the weights of the edges that connect vertices across the two subsets. In other words, we seek a graph bipartition that maximizes the weighted sum of ``cut'' edges.

To represent the graph's bipartitions, we associate a bitstring ${\boldsymbol{s} = (s_1, s_2, \dots, s_{N}) \in \{0, 1\}^N}$ with the vertices, where ${s_i = 0}$ or ${s_i = 1}$ indicates that vertex $i$ is assigned to subset $V_1$ or $V_2$, respectively, and ${N=|V|}$. Solving the signed Max-Cut problem equates to minimizing the objective function
\begin{equation}\label{eq:objective_function}
C(\boldsymbol{s}) = -\sum_{(i, j) \in E} J_{ij} (s_i \oplus s_j),
\end{equation}
which
calculates the negative total weight of the edges connecting vertices between the two subsets. Here, $\oplus$ denotes the addition modulo 2. By minimizing this objective function, we effectively maximize the original objective, i.e., the cut of the graph.
This type of objective function is not unique to the signed Max-Cut problem. Many combinatorial optimization problems can be cast into this form by mapping them onto a quadratic unconstrained binary optimization (QUBO) problem, often involving arbitrary real weights $J_{ij}$.

Let $C_\rmmin$ and $C_\rmmax$ represent the minimum and maximum value of the objective function, respectively, for a given graph. We define the normalized \emph{approximation ratio}
\begin{equation}\label{eq:classical_approx_ratio}
\bar r(\boldsymbol{s}) = \frac{C_\rmmax - C(\boldsymbol{s})}{C_\rmmax -C_\rmmin},
\end{equation}
which serves as a metric for evaluating the quality of a proposed solution. This ratio quantifies how close the objective function value $C(\boldsymbol{s})$ of a particular partition is to the optimal value $C_\rmmin$, where ${\bar r(\boldsymbol{s}) = 1}$ corresponds an optimal solution. For problem instances with non-negative weights ${J_{ij} \geq 1}$, the maximum value $C_\rmmax$ is null and Eq.~\eqref{eq:classical_approx_ratio} corresponds to the familiar expression ${\bar r(\boldsymbol{s})=C(\boldsymbol{s})/C_\rmmin}$.

Assigning a qubit to each vertex of the graph, the objective function can be encoded in the Hamiltonian
\begin{equation}\label{eq:Hp-qubo}
H_P = -\sum_{(i, j) \in E} J_{ij} (1-\sigma_i^z \sigma_j^z),
\end{equation}
where $\sigma_i^z$ represents the Pauli-$z$ operator acting on qubit $i$. The eigenvalue of $\sigma_i^z$ corresponds to the assignment of vertex $i$ to one of the two subsets, with eigenvalues $+1$ and $-1$ indicating membership in $V_1$ or $V_2$, respectively. In the computation basis, $H_P$ coincides with the objective function of Eq.~\eqref{eq:objective_function} up to global scaling factor $\frac{1}{2}$.
Thus, finding the ground state of $H_P$ is equivalent to finding the optimal partition that maximizes the sum of the signed edge weights between the two subsets, i.e., minimizes Eq.~\eqref{eq:objective_function}. 

We will analyze the signed Max-Cut problem on a complete graph, corresponding to solving the SK spin glass, in Sec.~\ref{subsec:SK-model}. In addition, we will provide results on 4-regular graphs in Sec.~\ref{sec:compiled_problems} and generalize the problem to hypergraphs in Appendix~\ref{sec:app:hypergraph}. For all our considerations, we use random weights ${J_{ij}\in\{-1, +1\}}$.

\subsection{Quantum Approximate Optimization}\label{sec:QAOA}
QAOA~\cite{Farhi2014, Blekos_PhysicsReports2024} is designed to find approximate solutions to combinatorial optimization problems, such as the signed Max-Cut problem. 

In {\vanilla} QAOA~\cite{Farhi2014}, a phase-separation unitary operator ${U_P(\gamma) = \eeuler^{-i\gamma H_P}}$ is applied alternately with a mixer unitary ${U_x(\beta) = \prod_{j=1}^{N} \eeuler^{-i\beta\sigma^x_j}}$ $p$-times to the initial state ${\ket{\Psi_0}=\ket{+}^{\otimes N}}$ on $N$ qubits. In this context, the integer number $p$ denotes the number of QAOA layers and is chosen before executing the algorithm. The resulting variational state 
\begin{equation}
    \label{eq:QAOA_circuit}
    \ket{\Psi(\bm{\beta},\bm{\gamma})} =  U_x(\beta_p) U_P(\gamma_p)\dots U_x(\beta_1) U_P(\gamma_1)\ket{\Psi_0}
\end{equation}
after $p$ layers depends on the $2p$ parameters ${\bm{\gamma}=(\gamma_1,\dots, \gamma_p)}$ and ${\bm{\beta}=(\beta_1,\dots, \beta_p)}$. 
A classical optimization routine is used within a quantum-classical feedback loop to find the parameters $(\bm{\beta^*}, \bm{\gamma^*})$ that minimize the objective function 
\begin{equation}
\calC(\bfbeta,\bfgamma)= \bra{\Psi(\bfbeta,\bfgamma)} H_P \ket{\Psi(\bfbeta,\bfgamma)}.
\end{equation}
The objective function is estimated by repeatedly preparing and measuring the variational state $ \ket{\Psi(\bm{\beta},\bm{\gamma})}$. 
For $p\to \infty$, the optimal value $\calC(\bfbeta^*,\bfgamma^*)$ converges monotonically to the ground state energy of $H_P$ and $\ket{\Psi(\bfbeta^*,\bfgamma^*)}$ converges to the optimization problem's solution~\cite{Farhi2014, Mbeng_Fazio_Santoro_arXiv2019}. For finite $p$, measuring $\ket{\Psi(\bfbeta^*,\bfgamma^*)}$ will provide approximate solutions. This implies that it is theoretically possible to find the exact solution for a problem for sufficiently large $p$. However, in practice, one often seeks an approximate solution and therefore limits $p$ to maintain a feasible optimization space. Then, the solutions obtained via {\vanilla} QAOA, have the approximation ratio 
\begin{equation}\label{eq:approx_ratio}
    r = \frac{C_\rmmax - \calC(\bfbeta^*,\bfgamma^*)}{C_\rmmax - C_\rmmin}.
\end{equation}

Due to the limited hardware connectivity of current quantum devices, a direct implementation of $U_P(\gamma)$ is usually not viable, as it would involve non-local entangling gates for all qubit pairs in $E$. While local interactions can be efficiently implemented using two CNOT gates and a single-qubit $z$-rotation, non-local interactions
typically necessitate a rerouting strategy~\cite{Hirata_QuantumInfoConf2011_LNN, Weidenfeller_quantum2022} involving additional SWAP gates (each comprising three CNOT gates) to establish the required interactions [see Fig.~\ref{fig:Figure1}(b)]. However, using SWAP networks results in undesired circuit depth overhead for {\vanilla} QAOA, which scales linearly with the number of qubits $N$ \cite{Weidenfeller_quantum2022, Kivlichan_Babbush_PRL2018, crooks_arXiv2018}. In addition, if SWAP gates are not efficiently arranged, their excessive use can lead to unnecessary noise and degrade the accuracy of the computation.

Several compilers~\cite{Sivarajah_IOPPublishing2021, Lingling_Proceedings2022, kotil_arXiv2023, Qiskit} provide algorithms to optimize the circuit depth and gate count of {\vanilla} QAOA on quantum hardware with linear~\cite{Harrigan_NaturePhysics2021, Hirata_QuantumInfoConf2011_LNN, crooks_arXiv2018} or square-grid connectivity~\cite{Weidenfeller_quantum2022}. For instance, when compiling complete graphs on a square-grid device, using linear SWAP networks is the optimal strategy, resulting in a CNOT depth~\footnote{We define the CNOT depth as the number of circuit layers comprising at least one CNOT gate.} of ${3N+4}$  and gate count of ${\frac{3}{2} N(N - 1)}$ per {\vanilla} QAOA layer~\cite{Weidenfeller_quantum2022, crooks_arXiv2018}.

\subsection{Parity Quantum Approximate Optimization}
\label{sec:ParityQAOA}
The parity transformation \cite{Lechner_2015, Ender2023} is an alternative solution for the limited connectivity of the quantum hardware by mapping multi-qubit entangling terms to single-qubit terms. The transformation replaces the original $N$ \emph{logical} qubits $\sigma_j^z$ with $K$ \emph{physical} qubits $\tilde{\sigma}_\nu^z$, such that each physical qubit represents the parity of a pair of logical qubits ${\sigma_{i_{\nu}}^z\sigma_{j_{\nu}}^z \to \tilde{\sigma}_{\nu}^z}$. Hereafter, we will mark physical states and operators with a tilde to distinguish them from logical ones.

Starting from  Eq.~\eqref{eq:Hp-qubo}, and applying the parity transformation results in the local Hamiltonian 
\begin{equation}\label{eq:local_field_hamiltonian}
    \tilde{H}_P= \sum_{\nu=1}^K \tilde{J}_{\nu}\tilde{\sigma}_\nu^z
\end{equation}
for the physical qubits,
where ${\tilde{J}_{\nu}=J_{i_{\nu}j_{\nu}}}$  with ${\nu=1, 2, ...K}$ are local fields for the physical qubits. In particular, the problem of finding the ground state of the original Hamiltonian $H_P$ is equivalent to the problem of finding the minimum energy state of the physical Hamiltonian ${\tilde{H}_P}$ with ${L=K-N+D}$ independent constraints, where $D$ denotes the number of logical degeneracies of $H_P$ (see Ref.~\cite{Ender2023} for details). The role of these constraints is to enforce that all considered configurations of physical qubits correspond to a unique configuration of logical qubits. Ref.~\cite{Lechner_2015} shows that this condition can be attained by arranging the qubits on a square-grid planar geometry and choosing the set of $L$ local constraints to be
\begin{equation}\label{eq:parity-constraints}
    \tilde{H}_{\squaredots,l}\ket{\tilde{\psi}} = \ket{\tilde{\psi}} 
    \qquad \mbox{for}\;\; l=1,2, \dots,L\,,
\end{equation}
such that ${\tilde H_{\squaredots,l}=\Tilde{\sigma}_{(l,1)}^z\Tilde{\sigma}_{(l,2)}^z\Tilde{\sigma}_{(l,3)}^z[\Tilde{\sigma}_{(l,4)}^z]}$ are 
 local four- and three-qubit plaquette terms on the grid. The correct choice for these constraint terms depends on the original optimization problem. In particular, complete problem graphs are known to require ${K=\frac{N(N-1)}{2}}$ physical qubits and ${L=\frac{(N-1)(N-2)}{2}}$ constraints arranged in a Lechner-Hauke-Zoller geometry~[see Fig.~\ref{fig:Figure1}(b)]\cite{Lechner_2015}. Recently, Ref.~\cite{Ender2023} extended the parity transformation to Polynomial Unconstrained Boolean Optimization (PUBO) problems by encoding the parity of multiple logical qubits in a single physical qubit and designing efficient algorithms to identify the correct geometry for the physical qubits and the plaquette terms~\cite{parityos}.


In {\parity} QAOA \cite{Lechner2018,Weidinger2023}, two phase-separation unitary operators ${\tilde{U}_z(\gamma) = \eeuler^{-i\gamma\tilde{H}_p}}$ and ${\tilde{U}_{\squaredots}(\Omega) = \prod_{l=1}^{L}\eeuler^{-i\Omega \tilde{H}_{\squaredots,l}}}$ are applied alternately with a mixer unitary ${\tilde{U}_x(\beta) = \prod_{\nu=1}^{K} \eeuler^{-i\beta\Tilde{\sigma}_j^x}}$ $p$-times to the initial state ${\ket{\tilde \Psi_0}=\ket{+}^{\otimes K}}$ on $K$ qubits.
Adding another set of variational parameters  ${\bfOmega=(\Omega_1, \dotsm \Omega_p)}$ for the constraint unitaries $\tilde{U}_{\squaredots}$ results in better performance than merging it with the main cost Hamiltonian into a single phase operator, as demonstrated in \cite{Lechner2018}.
The resulting variational state for  {\parity} QAOA is
\begin{equation}
\begin{split}
\ket{\tilde{\Psi}(\bfbeta,\bfOmega,\bfgamma)} &=  \tilde{U}_x(\beta_p) \tilde{U}_{\squaredots}(\Omega_p) \tilde{U}_z(\gamma_p)\dots\\
&\hspace{1.2cm}\dots\tilde{U}_x(\beta_1) \tilde{U}_{\squaredots}(\Omega_1) \tilde{U}_z(\gamma_1)\ket{+}^{\otimes K},
\end{split}
\label{eq:Parity_circuit}
\end{equation}
Ref.~\cite{Lechner2018} showed that, for $p\to \infty$, this state can represent the minimum energy state of $\tilde{H}_P$ with the constrains in Eq.~\eqref{eq:parity-constraints}, which corresponds to the original optimization problem's solution. 

{\Parity} QAOA is particularly suited for quantum hardware with planar square-grid connectivity. Therefore, we consider square-lattice hardware platforms for this study. Indeed, implementing ${\tilde{U}_z(\gamma)}$ and ${\tilde{U}_x(\beta)}$ only requires composing 
single-qubit gates, with no hardware connectivity requirements. On the other hand, the four-qubit (three-qubit) plaquette gates ${\eeuler^{-i\Omega \tilde{H}_{\squaredots,l}}}$ can be implemented with a chain of six (four) CNOT gates and a single-qubit $z$-rotation [see Fig.~\ref{fig:Figure1}(c)]. All constraint terms can be implemented in parallel, resulting in a constant circuit depth per QAOA layer~\cite{Lechner2018, Unger_Lechner_arXiv2023}. Specifically, for a QUBO problem on a complete graph, the CNOT depth is $10$ (independent of $N$), and the gate count per QAOA layer is ${2(N - 2)(N - 3)}$ on a square grid~\cite{Unger_Lechner_arXiv2023}. The CNOT depth and gate count per layer for other graphs can be obtained by following the procedure in Ref.~\cite{Unger_Lechner_arXiv2023}.


To run {\parity} QAOA, we also need to specify a measurable objective function that is minimized when the optimization problem is solved. The original {\parity} QAOA objective function, introduced in Ref.~\cite{Lechner2018}, uses penalty terms to enforce the plaquette constraints of Eq.~\eqref{eq:parity-constraints}. 
However, as pointed out in Ref.~\cite{Weidinger2023}, the algorithm performs better
when the projective measurements of the physical $K$-qubit variational state are explicitly decoded into classical configurations of the $N$ logical variables.

Following Ref.~\cite{Weidinger2023}, we thus employ a decoding strategy that assigns values ${\bfq=(q_1,\dots,q_N)\in\{0, 1\}^N}$ to $N$ logical qubits based on a given readout measurement of a subset of the $K$ physical qubit measurement outcome ${\tilde{\bfq}=(\tilde{q}_1,\dots,\tilde{q}_K)}$. We use the term `readout basis' to refer to these subsets of physical qubits.  
On a graph with a single logical degeneracy ($D=1$), the readout basis of the physical qubits represents the edges of a spanning tree $t$ in the logical graph; that is, the corresponding edges form a tree that includes all nodes of the problem graph. 
An example of a single tree is given in Fig.~\ref{fig:Figure1}(c) on the right.
A spanning tree contains $N-1$ edges, hence, $N-1$ parity qubits must be in the readout basis.
The measurement outcome of a parity qubit describes the parity between two logical values. 
A readout of $0$ means that the corresponding logical qubit values are identical (e.g., $0$ and $0$) and, conversely, if the readout is $1$ the logical qubits values are different (e.g., $0$ and $1$). For each spanning tree $t$, applying these rules to all $N-1$ readout variables, we can determine a decoded logical state ${\bfq}^{(t)}$ (up to a global spin flip). Different spanning trees can lead to different logical values.
For general graphs and hypergraphs, readout bases have $N-D$ physical qubits and are not necessarily spanning trees~\cite{Ender2023}. 

We define the projector 
\begin{equation}
    P_{\bfq^{(t)}} = \ket{\bfq^{(t)}}\bra{\bfq^{(t)}}
\end{equation}
onto the decoded logical state $\bfq^{(t)}$. A 
 decoding map $D_t$ can then be defined as
\begin{equation}
D_{t}[\tilde{\Psi}(\bfbeta,\bfOmega,\bfgamma)]=\!\!\!\sum_{\scriptscriptstyle\tilde{\bfq}\in \{0,1\}^K} \!\!\! |\braket{\tilde{\bfq}|\tilde{\Psi}(\bfbeta,\bfOmega,\bfgamma)}|^2 P_{\bfq^{(t)}}.
\end{equation}
In this description, the measurement outcome $\tilde{\bfq}$, which appears with probability $|\braket{\tilde{\bfq}|\tilde{\Psi}(\bfbeta,\bfOmega,\bfgamma)}|^2$, is translated to a logical configuration $\bfq^{(t)}$.

Considering a set ${T=\{t_1, t_2,\dots t_M\}}$ of $M$ spanning trees used for decoding, we can define the objective function as
\begin{align}
     \tilde{\calC}^{\mathrm{mean}}(\bfbeta, \boldsymbol{\Omega},\bfgamma)
     =\frac{1}{M} \sum_{m=1}^{M} & \Tr  \left( H_P \,D_{t_m}[
     \tilde{\Psi}(\bfbeta,\bfOmega,\bfgamma) ]\right)\label{eq:C_parity_mean},
\end{align}
which corresponds to the mean energy over the different spanning tree readouts $\bfq^{(t_i)}$ with ${i=1, \dots, M}$. An alternative way to define the objective function is to consider the best (lowest) energy of the $M$ spanning tree readouts 
\begin{align}
     \tilde{\calC}^{\mathrm{best}}(\bfbeta,\bfOmega,\bfgamma)
     =\underset{t \in T}{\min} & \Tr  \left( H_P \,D_{t}[
     \tilde{\Psi}(\bfbeta,\bfOmega,\bfgamma) ]\right)\label{eq:C_parity_min}.
\end{align}
Then, the approximation ratio's expected value for the solutions obtained via {\parity} QAOA is
\begin{equation}\label{eq:approx_ratio_parity}
    r = \frac{C_\rmmax - \tilde{\calC}(\bfbeta^*,\bfOmega^*, \bfgamma^*)}{C_\rmmax - C_\rmmin},
\end{equation}
where  $\tilde{\calC}(\bfbeta^*,\bfOmega^*, \bfgamma^*)$ can either be the mean $\tilde{\calC}^{\mathrm{mean}}$ or the best spanning tree energy $\tilde{\calC}^{\mathrm{best}}$, as defined in Eq.~\eqref{eq:C_parity_mean} and Eq.~\eqref{eq:C_parity_min}.

For a complete graph, $N^{N-2}$ spanning trees exist. Ref.~\cite{Weidinger2023} showed that the performance of Parity QAOA increases the more spanning trees are used but does not change significantly for a particular choice of trees. 
Following Ref.~\cite{Weidinger2023}, we limit the classical decoding overhead by considering a specific set of ${M=N}$ spanning trees, namely the \emph{logical lines} in the parity encoding, denoted by ${T^{(l)} = \{t_1^{(l)}, t_2^{(l)}, \dots t_N^{(l)}\}}$. The logical line $i$, corresponding to spanning tree $t_i^{(l)}$, includes all parity qubits that contain information about the logical qubit $i$. 
(i.e., all qubits ${\tilde{\sigma}_{\nu}}$ that encode the parity of the logical qubits $\sigma_{i}$ and $\sigma_{j}$ where $j=1,\dots, N$, excluding $j=i$). This means that for a complete graph with $N$ nodes there are exactly $N$ logical lines.
Figure~\ref{fig:Figure1}(c) shows a spanning tree readout of $t_5^{(l)}$.

\subsection{Clifford  pre-optimization}\label{sec:clifford_circuit}

The term Clifford circuit pre-optimization describes a classical strategy to find high-quality (initial) parameters before running a VQA (e.g., QAOA) on a quantum device~\cite{MunosArias_Blais_PRR2024, Cheng_Kim_arXiv2022}. This section introduces Clifford gates and their use in QAOA  pre-optimization. More information on the topic can be found in Refs.~\cite{Aaronson2004,Cheng_Kim_arXiv2022,MunosArias_Blais_PRR2024}.

The defining property of Clifford gates is mapping Pauli matrices and their tensor products into themselves by conjugation. More precisely, let $\boldsymbol{\mathcal{P}}_N$ be the $N$-qubit Pauli unitary group generated by the single qubit operations ${\{\sigma_j^x, \sigma_j^y,\sigma_j^z\}_{j=1,\dots,N}}$. Then, $N$-qubit Clifford gates are unitary transformations $G$ such that ${G^\dagger X G \in \boldsymbol{\mathcal{P}}_N}$ for all ${X \in \boldsymbol{\mathcal{P}}_N}$. 
Their relevance in QAOA follows from  the Gottesman-Knill theorem~\cite{Gottesman_phDThesis, Gottesman1998,Aaronson2004}, which ensures that Clifford gates can be simulated efficiently (in polynomial time)  on classical computers. 

For the signed Max-Cut problem class, the {\vanilla} QAOA unitaries $U_\kappa(\theta)$ with ${\kappa=x, P}$
  and the {\parity} QAOA unitaries $\tilde{U}_\kappa(\theta)$ with ${\kappa=x, {\squaredots}, P}$ 
are Clifford gates if ${\theta \in A=\{-\frac{\pi}{4}, 0, \frac{\pi}{4}, \frac{\pi}{2}\}}$~\footnote
{
    This statement follows from the observation that for any non-trivial hermitian Pauli string $X \in\boldsymbol{P}_N$ the exponential $\eeuler^{i\theta X}$ is a Clifford gate if and only if $\theta$ is an integer multiple of $\pi/4$. Moreover, since $\eeuler^{-i\theta X}Y\eeuler^{i\theta X}$ has a period of $2\pi$, we only need to consider the values ${\theta \in \{-\frac{\pi}{4}, 0, \frac{\pi}{4}, \frac{\pi}{2}\}}$.
}.
Therefore, when ${\bfbeta, \bfOmega, \bfgamma\in A^{p}}$, we can use the Gottesman-Knill theorem to efficiently sample the {\vanilla} and {\parity} QAOA objective functions on a classical computer. The Clifford circuit pre-optimization is a search aimed at (locally or globally) minimizing the QAOA objective function, with the constraint ${\bfbeta, \bfOmega, \bfgamma\in A^{p}}$. This approach's main advantage is that the search algorithm can run efficiently on a classical computer with a query cost polynomial in the problem size $N$.

Let ${\bfbeta^{(0)}, \bfOmega^{(0)}, \bfgamma^{(0)}\in A^{p}}$ be the optimal parameters found by the Clifford circuit pre-optimization. These parameters can be used as initial values for  QAOA. In this case, QAOA's approximation ratio is bounded by $r>r^{(0)}$, where $r^{(0)}$ is the approximation ratio associated with the found optimal Clifford circuit:
\begin{equation}
    r^{(0)} = \frac{C_\rmmax - \tilde{\calC}(\bfbeta^{(0)}, \bfOmega^{(0)} \bfgamma^{(0)}) }{C_\rmmax - C_\rmmin}\,. 
\end{equation}
The value of $r^{(0)}$ thus provides a classical lower bound to the accuracy of the QAOA algorithm. This lower bound is particularly useful for predicting QAOA's performance on large problem sizes, which cannot be simulated on classical computers.

\setlength{\tabcolsep}{5pt}
\renewcommand{\arraystretch}{1.5}
\begin{table}
\centering
\begin{tabular}{c | c | c | c} 
  $l$ & $\gamma_{l}$ & $\Omega_{l
  }$ &  $\beta_{l}$  \\ [0.5ex] 
  \hline\hline
  $1,\dots\frac{p-2}{2}$ & $0$ & $\frac{1}{4}$ & $\frac{1}{4}$\\
  \hline
  $\frac{p}{2}$ & $0$ & $\frac{1}{4}$ & $\frac{1}{2}$ \\
  \hline
    $\frac{p+2}{2},\dots(p-1)$ & $0$/$\frac{1}{2}$ & $\pm\frac{1}{4}$ & $\frac{1}{4}$ \\
    \hline
    $p$ & $\pm\frac{1}{4}$ & $\pm\frac{1}{4}$ & $\frac{1}{4}$ \\
 [1ex] 
 \hline
\end{tabular}
\caption{Classical parameter vectors for even QAOA layers ${p>2}$. Values for arbitrary $p$ can be found in Appendix~\ref{sec:app:classical_clifford_angles}. The table shows the index values $l=1,\dots, p$, for the parameters $(\gamma_{l}, \Omega_{l}, \beta_{l})$ (which are given in units of $\pi$).}
\label{tab:Parameters_even}
\end{table}

\section{Results}

We present data from numerical simulations that aim to compare {\vanilla} and {\parity} QAOA's performances on the signed Max-Cut problem. We focus on complete and 4-regular graphs, using the circuits' CNOT depth and CNOT counts to quantify the hardware resources required to run the two algorithms on a square-grid architecture. We conclude the section by introducing a recursive variant {\parity} QAOA variant and comparing it with the conventional RQAOA approach.

\subsection{Performance lower bounds for parity QAOA}\label{sec:ParityLowerBound}

The quadratic qubit overhead introduced by parity QAOA makes simulating the algorithm on large graphs (${N>20}$) a challenging task. Therefore, we use Clifford pre-optimization to establish lower bounds on {\parity} QAOA's performance, which we will later compare with {\vanilla} QAOA's results (see Sec.~\ref{sec:QAOA_comparison}).

We can restrict the optimal parameter search to the subset ${\Aclassical\subset A^{p}\times A^{p}\times A^{p}}$ to further reduce the Clifford pre-optimization's running time (for a detailed discussion see Appendix~\ref{sec:app:classical_clifford_angles}). Here, we chose $\Aclassical$ such that for 
${(\bfgamma, \bfOmega, \bfbeta) \in \Aclassical}$ 
the prepared parity QAOA state $\ket{\tilde{\Psi}(\bfbeta,\bfOmega,\bfgamma)}  $ is an element of the computational basis ${\{\ket{\tilde{q}_{1}, \dots, \tilde{q}_{K}}\,\,|\,\, \tilde q_i = 0, 1\}}$. 
The elements of $\Aclassical$, for the signed Max-Cut problem, are explicitly given in Tab.~\ref{tab:Parameters_even}.
We refer to them as \emph{classical parameter vectors} and to the corresponding states as \emph{classical states} in the following. 
The number of classical parameter vectors increases exponentially with the number of layers $p$, following the relation ${|\Aclassical|=2^{p+1}}$.
We refer to Appendix~\ref{sec:app:classical_clifford_angles} for more details about these classical parameter vectors. 

\begin{figure}
    \centering
    \includegraphics[width=\columnwidth]{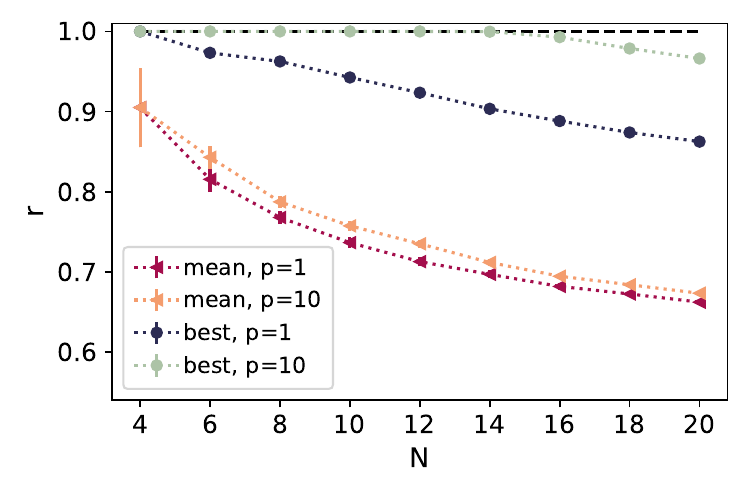}
    \caption{The approximation ratio $r$ as a function of the logical qubit number $N$ for the objective functions $\tilde{\calC}^{\mathrm{mean}}$ [red and orange line, Eq.~\eqref{eq:C_parity_mean}], and $\tilde{\calC}^{\mathrm{best}}$ [blue and green line, Eq.~\eqref{eq:C_parity_min}], for the decoded classical states with ${M=N}$ spanning trees. Data points represent averages over 50, 400, and 800 instances for ${N=4}$, $6$ and ${N>6}$, respectively, with error bars showing twice the standard error of the mean.}
    \label{fig:Decoding_angles}
\end{figure}

\begin{figure*}
    \centering 
    \includegraphics[scale=0.5]{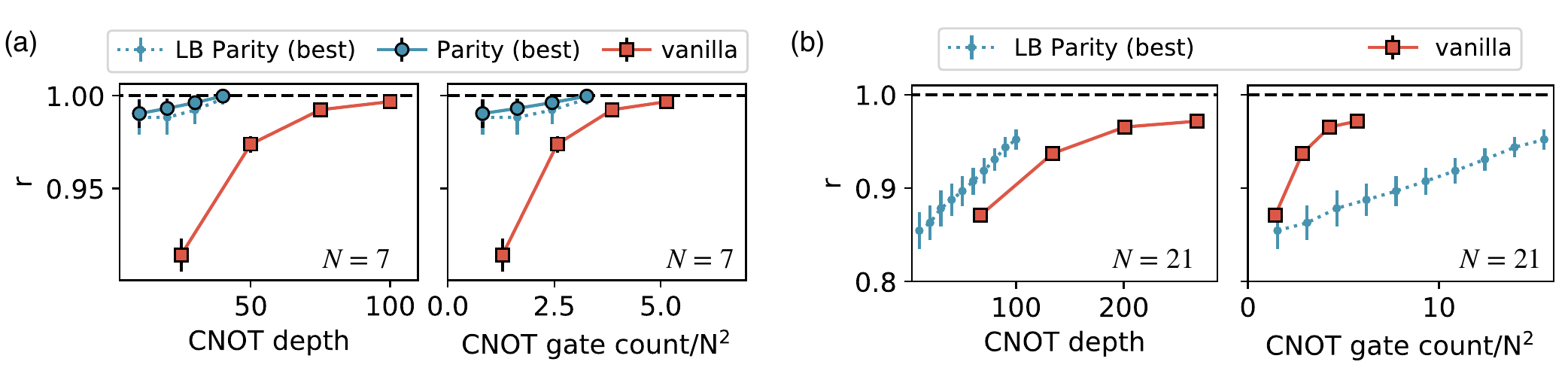}
    \caption{Approximation ratio $r$ for {\vanilla} and {\parity} QAOA as a function of CNOT depth and CNOT gate count. In each line, the left-most data point represents ${p=1}$, the second left-most data point represents ${p=2}$, and so forth.
    (a) Results for ${N=7}$ logical qubits (${K=21}$ physical qubits). Dotted lines indicate the lower bound (LB) results obtained by decoding classical states. Classical optimization parameters are: ${n_\text{init}=80}$, ${n_\text{mc}=400/1000/1500/2000}$ for ${p=1/2/3/4}$.
    Results represent the average over 100 simulated instances.
    (b) Results for ${N=21}$ (${K=210}$ physical qubits). Due to the system size, simulation of {\parity} QAOA is not feasible; therefore, only the LB is shown. Classical optimization parameters for {\vanilla} QAOA are ${n_\text{init}=50}$ and ${n_\text{mc}=200/500/800/1000}$ for ${p=1/2/3/4}$. For the LB of {\parity} QAOA, data is shown for higher $p$ than for {\vanilla} QAOA due to the feasibility of the simulations.
    Results represent the average over 40 simulated instances.
    In both plots, the error bars represent twice the standard error of the mean.
   }
    \label{fig:N7_optimization}
\end{figure*}

We now briefly describe the procedure we used to compute the lower bounds on the approximation ratio for {\parity} QAOA applied on the signed Max-Cut problem on a complete graph. 
For each classical parameter vector, we use the python library \emph{stim}~\cite{gidney_quantum2021} to efficiently simulate the {\parity} QAOA circuit and compute the output state.  
As each circuit returns a classical state, which is an element of the computational basis, no sampling is required.
We, then, decode the hereby obtained classical states using the spanning trees $T^{(l)}$ as described in Sec.~\ref{sec:ParityQAOA}. 
After evaluating the objective function for the resulting state according to Eq.~\eqref{eq:C_parity_mean} or  Eq.~\eqref{eq:C_parity_min}, we use the lowest measured energy of the $2^{p+1}$ states to compute the lower bound on the approximation ratio. The complexity of the whole procedure scales exponentially with the QAOA layers $p$.

Figure~\ref{fig:Decoding_angles} shows the results for the approximation ratio, as defined in Eq.~\eqref{eq:approx_ratio_parity}, as a function of the system size $N$, averaged over randomly generated instances (see Figure caption). An approximation ratio of ${r=1}$, marked with a black dashed line, signifies that the exact solution (configuration with the lowest objective function) is found.

The results indicate that choosing the best spanning tree energy $\tilde{\calC}^{\mathrm{best}}$ (best, ${p=1}$) yields a significantly higher approximation ratio than taking the mean $\tilde{\calC}^{\mathrm{mean}}$ (mean, ${p=1}$). In fact, using the objective function given by Eq.~\eqref{eq:C_parity_min} and ${p=1}$ QAOA layer yields an approximation ratio of ${r=1.0}$ for ${N=4}$, indicating that all problem instances were solved. For ${p=10}$ and $\tilde{\calC}^{\mathrm{best}}$, the classical parameter vectors generate the solution for every instance up to ${N=12}$. In contrast to that, using the mean energy $\tilde{\calC}^{\mathrm{mean}}$ does not show significant improvement when increasing the number of layers from ${p=1}$ to ${p=10}$.

In Appendix~\ref{sec:app:decoding_classical_states} we elaborate more on the success rate of decoding classical states. Additionally, Appendix~\ref{sec:app:optimization_SKmodel} presents results for QAOA simulations for both the best $\tilde{\calC}^{\mathrm{best}}$ and mean $\tilde{\calC}^{\mathrm{mean}}$ objective function. In the following sections we choose the best spanning tree energy $\tilde{\calC}^{\mathrm{best}}$ as the objective function.

We note that we did not attempt a similar analysis for {\vanilla} QAOA, as Ref.~\cite{Cheng_Kim_arXiv2022} observed that Clifford pre-optimization for Max-Cut problems is not a helpful strategy for {\vanilla} QAOA. 
Nevertheless, there are cases where Clifford gates play a significant role in {\vanilla} QAOA~\cite{MunosArias_Blais_PRR2024, wilkie_Herrman_arXiv2024, Vijendran_IOPscience2024}.

\subsection{QAOA simulations}\label{sec:QAOA_comparison}
Next, we investigate the performance of {\vanilla} and {\parity} QAOA with respect to CNOT gate count and circuit depth for the signed Max-Cut problem on a complete and a 4-regular graph. 
{\Vanilla} QAOA simulations are done with $K/N$ copies to equalize the number of qubits required in both embeddings. QAOA is executed on all copies in parallel, where, after measurement of all qubits, the lowest energy of all copies is considered to evaluate the objective value.
However, we do not multiply the CNOT depth or gate count by the $K/N$ factor; instead, we compare the hardware requirements to run the individual QAOA circuits.
The initial parameters for the classical optimization for both QAOA methods are chosen randomly from the interval $[-\frac{\pi}{2}, \frac{\pi}{2})$. In particular, our QAOA simulations do not use the Clifford pre-optimization described in Sec.~\ref{sec:clifford_circuit} to enhance the algorithms' performance.
More details are provided in Appendix~\ref{sec:app:simulation_details}. 
The QAOA circuit is simulated with the Python library \emph{qiskit}~\cite{Qiskit}. Furthermore, we compare the QAOA simulation results with the lower bounds, which were computed in the last section.

\subsubsection{Complete graph \label{subsec:SK-model}}

We start by analyzing the performance of {\parity} QAOA for signed Max-Cut on a complete graph, i.e., the SK spin glass.
Figure~\ref{fig:N7_optimization}(a) shows the average approximation ratio for 100 random instances with ${N=7}$ nodes, for both {\vanilla} and {\parity} QAOA. The approximation ratio is evaluated  with respect to the CNOT depth on the left and the CNOT gate count on the right.
The circuit depth and the gate count for $p$ QAOA layers in both approaches were obtained from the numbers for a single layer discussed in Secs. \ref{sec:QAOA} and \ref{sec:ParityQAOA}.

Each data point corresponds to a specific QAOA depth $p$: the left-most point represents one layer ${(p=1)}$, the second left-most point represents two layers ${(p=2)}$, and so on.
The blue dotted line indicates the lower bound (LB) presented in Sec.~\ref{sec:ParityLowerBound}.
When QAOA is applied, the approximation ratio exceeds the lower bound. 

The data points for the lower bound and the actual optimization results lie close to each other because the decoding yields the ground state for most instances, e.g., for $96$ out of $100$ for ${p=3}$. We refer to these instances as \emph{trivial} instances. In contrast, instances not solvable by decoding the classical states alone are termed \emph{non-trivial} instances. Increasing the number of QAOA layers to ${p=4}$ reduces the number of non-trivial instances to one non-trivial instance within this sample size.
For cases where decoding classical states does not produce the ground state, the application of QAOA further improves the approximation ratio.
Simulations of {\parity} QAOA for non-trivial instances only are presented in Appendix~\ref{sec:app:Optimization}.

In terms of both, CNOT depth and CNOT gate count, {\parity} QAOA achieves a higher approximation ratio using the same quantum resources for ${N=7}$. However, this behavior is no longer observed when $N$ increases, as shown in  Fig.~\ref{fig:N7_optimization}(b), which presents results for ${N=21}$. Simulations for this scenario, involving ${K=210}$ physical qubits, can not be simulated on a classical computer. Thus, only the lower bound of the approximation ratio is reported. While the {\parity} QAOA lower bound shows an advantage in terms of CNOT depth (left), {\vanilla} QAOA is superior in terms of CNOT gate count (right). 
This is expected, as the depth for one layer of {\parity} QAOA is independent of the system size, while it scales linearly with $N$ for {\vanilla} QAOA. In leading order, the CNOT gate count scales with $2N^2$ for {\parity} QAOA and $\frac{3}{2}N^2$ for {\vanilla} QAOA. 

Appendix~\ref{sec:app:Comparison} provides a more detailed discussion on the differences in implementation, qubit overhead and required post-processing resources of the two QAOA approaches. A comparison of explicit numbers for different metrics is given in Tab.~\ref{tab:app:Comparison}.

\begin{figure}
\includegraphics[width=\columnwidth]{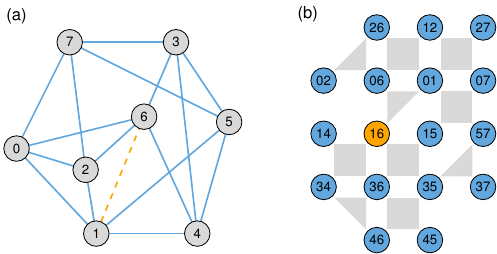}
    \caption{(a) 4-regular graph in the logical representation. The interaction corresponding to the physical ancilla qubit (16) is shown as a yellow dashed line. (b) Compiled parity graph obtained from the 4-regular graph shown in (a), with the ancilla qubit (16) colored yellow. }
    \label{fig:4-regular_graph}
\end{figure}

\subsubsection{4-regular graph}\label{sec:compiled_problems} 
The parity transformation can also be done for non-complete graphs \cite{Ender2023}. An arbitrary example, with lower connectivity and small enough to classically simulate, is the 4-regular graph with ${N=8}$ nodes shown in Fig.~\ref{fig:4-regular_graph}.
The parity graph was obtained by the ParityOS compiler~\cite{Ender2023, parityos}, which introduced the ancilla qubit $(16)$ to complete the mapping.
To compare the performance of {\parity} QAOA on such \textit{compiled graphs}, we again consider 100 random choices of the weights ${J_{ij}}$.

For compiled graphs not all edges are present and, therefore, the ${M=N}$ logical lines as described before do not exist. However, like in Ref~\cite{Weidinger2023}, we choose a random set of spanning trees. Regarding the number of trees, we could, in principle, consider a set $T=\{t_1, t_2, ..., t_8\}$ of ${M=N=8}$ spanning trees, that uses the same number of spanning trees as the complete graph.
Considering the fact that the compiled graph has less physical qubit overhead than the complete graph one can also choose a set with fewer spanning trees, such that the  average number of spanning trees per physical qubit $D_M$ equals the number of logical lines for the complete graph. One can calculate $D_M$ as follows.
A spanning tree covers ${N-1}$ parity qubits, and with $M$ spanning trees over $K$ parity qubits we obtain the relation ${D_M=M(N-1)/K}$.
In the complete graph, we have ${D_M=2}$ for ${M=N}$, which means that each parity qubit is contained in 2 spanning trees.
To obtain a similar value in the 4-regular graph, whose parity mapping is depicted in Fig.~\ref{fig:4-regular_graph}(b), we require ${M=5}$ spanning trees, $T=\{t_1, \dots, t_5\}$, yielding ${D_M=2.06}$.
In Appendix~\ref{sec:app:4regular_optimization}, we show a comparison of the two choices of spanning trees (with $M=8$ and $M=5$, see Fig.~\ref{fig:app:SuccessRate_allJij} (b)) and give the explicit spanning trees we used for decoding. In the main text, we will only consider the set with $M=5$ spanning trees. 

\begin{figure}
    \centering
    \includegraphics[width=\columnwidth]{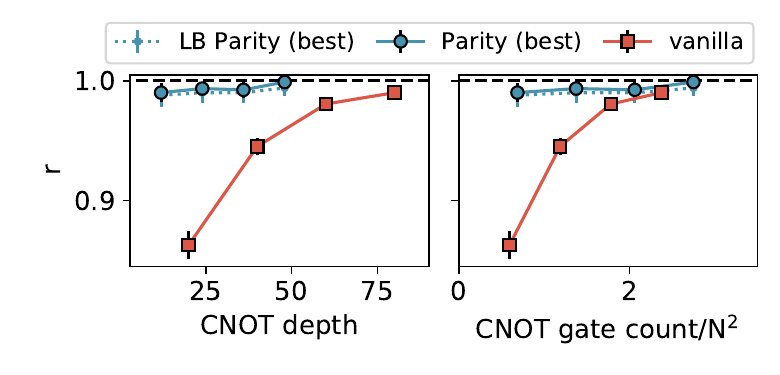}
    \caption{Approximation ratio $r$ for {\vanilla} and {\parity} QAOA as a function of CNOT depth and CNOT gate count for a 4-regular graph with ${N=8}$ nodes and ${M=5}$. In each line, the left-most data point represents ${p=1}$, the second left-most data point represents ${p=2}$, and so forth. The dotted lines show the LB of the approximation ratio.
    Classical optimization parameters are: ${n_\text{init}=80}$, ${n_\text{mc}=400/1000/1500/2000}$ for ${p=1/2/3/4}$.
    Results show the average over 100 simulated instances.
   The error bars represent twice the standard error of the mean.
   }
    \label{fig:N8_optimization_4reg}
\end{figure}

Following the constraint implementation procedure in Ref.~\cite{Unger_Lechner_arXiv2023}, the {\parity} QAOA circuit has a CNOT depth of $12$ and a CNOT-gate count of $44$. 
The circuit for the {\vanilla} QAOA was transpiled by $\text{t}|\text{ket}\rangle$~\cite{Sivarajah_IOPPublishing2021} and then further optimized. The resulting circuit has a CNOT depth of $20$ and a CNOT gate count of $38$. 
Results for QAOA simulations for 100 problem instances are shown in Fig.~\ref{fig:N8_optimization_4reg}. Again, due to the high LB for the success rate of {\parity} QAOA, the approximation ratio reaches ${r>0.97}$. There are 6/5/3/3 non-trivial instances for ${p=1/2/3/4}$ for 100 simulated instances. 
These results are in accordance with the data obtained for the complete graph.
Simulations for exclusively non-trivial instances are discussed in Appendix~\ref{sec:app:4regular_optimization}. 

\subsection{Recursive QAOA simulations}\label{sec:recursive_qaoa}

QAOA's performance and limitations on regular graphs were studied in Refs.~\cite{Bravyi_Tang_PRL2020, Farhi_Gutmann_arXiv2020,Mbeng_Fazio_Santoro_arXiv2019,Wybo_Leib_Quantum2024}. The authors showed that QAOA needs to ``see the whole graphs'' to avoid the obstacles arising from the algorithm's locality. In particular, Ref.~\cite{Farhi_Gutmann_arXiv2020} proved that approximating the solutions of typical random regular graphs requires scaling the number of QAOA layers with the graph size at least logarithmically. Ref.~\cite{Wybo_Leib_Quantum2024} extended these results to the {\parity} architecture, showing that approximating the solutions of complete graphs with {\parity} QAOA requires a number of layers that scales at least as $\mathcal{O}(\sqrt{N})$.

Ref.~\cite{Bravyi_Tang_PRL2020} proposed a recursive QAOA (RQAOA) to solve Max-Cut problems to overcome the performance limitations induced by locality. Starting from the original $N$-qubit problem Hamiltonian $H^{(N)}_p=H_p$, the RQAOA iteratively reduces the problem's size by a procedure called \emph{correlation rounding}~\cite{Bravyi_Tang_PRL2020, Bravyi_Tang_Quantum2022}. Given an $N'$-qubit Hamiltonian $H^{(N')}_p$, a single variable elimination via correlation rounding works as follows: First we use the QAOA subroutine to compute the state $\ket{\Psi^{(N')}(\bfbeta^*,\bfgamma^*)}$ that best approximates the ground state of $H^{(N')}_p$. Then we use repeated measurements to estimate the correlation matrix 
\begin{align}
    W^{(N')}_{jk}&= \bra{\Psi^{(N')}(\bfbeta^*,\bfgamma^*)} \sigma_j^z \sigma_k^z \ket{\Psi^{(N')}(\bfbeta^*,\bfgamma^*)}\;.\label{eqn:correlation-matrix}
\end{align}
Next, we compute the edge with the largest absolute correlations
\begin{align}
    \left(\bar{j}, \bar{k}\right)&= \argmax_{(j, k)\in E} \left|W^{(N')}_{jk}\right|.
\end{align}
Finally, we reduce the problem size by the fixing rule 
\begin{align}
    \sigma^z_{\bar{j}} = \mathrm{sign}\left( W^{(N')}_{\bar{j}, \bar{k}}\right)\sigma^z_{\bar{k}} \label{eqn:fixing-rule}
\end{align}
substituting into $H^{(N')}$. The substitution results in an new ${(N'-1)}$-qubit problem Hamiltonian $H^{(N'-1)}$.
The variable-elimination process is repeated on the increasingly
smaller graph until the problem becomes trivial to solve.
Then, the sequence of fixing rules is reversed to obtain an approximate solution for the original problem.

The non-local operations involved in RQAOA's correlation rounding can overcome QAOA's performance limitations~\cite{Bravyi_Tang_Quantum2022}. In particular, several numerical and theoretical studies~\cite{Bravyi_Tang_Quantum2022,Finzgar_KatzgraberPRX2024, Brady_Hadfield_arXiv2023} suggest that using RQAOA leads to finding better approximate solutions with performances that are comparable to state-of-the-art general-purpose classical solvers. In the following, we introduce a parity-based implementation of RQAOA by making suitable changes to the original RQAOA algorithm. We use the terms {\vanilla} RQAOA and {\parity} RQAOA to distinguish between the original and the new parity-based RQAOA versions.

To implement {\parity} RQAOA,
 we replace the correlation matrix of Eq.~\eqref{eqn:correlation-matrix} with 
 \begin{align}
    W^{(N')}_{jk}&= \frac{1}{M} \sum_{m=1}^{M}  \Tr  \left( \sigma_j^z \sigma_k^z \,D_{t_m}[
     \tilde{\rho}^{(N')}(\bfbeta^*,\bfOmega^*,\bfgamma^*) ]\right)\label{eq:C_parity_mean_recursive}\;.
\end{align}
Then, we use correlation rounding to iteratively eliminate logical variables from the original problem Eq.~\eqref{eqn:fixing-rule}.

\begin{figure}
    \centering
    \includegraphics[width=\columnwidth]{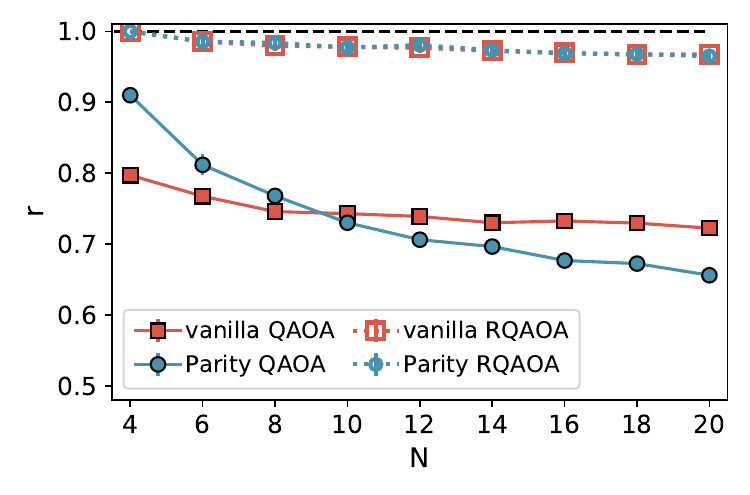}
    \caption{
    Approximation ratio $r$ for single layer RQAOA ($p=1$) 
    on complete graphs. Vanilla RQAOA (red empty squares) and parity RQAOA (blue empty circles) approximation ratios are shown as a function of the graph size $N$. The filled symbols correspond to the single-layer QAOA simulations on the same problem graph. The objective function $\tilde{\calC}^{\mathrm{mean}}$ [Eq.~\eqref{eq:C_parity_mean}] was used for the {\parity} architecture. Data points represent averages over 100 simulated instances. The error bars are twice the standard error of the mean.}
    \label{fig:rqaoa}
\end{figure}

Although analytical results on {\parity} RQAOA's performance are not available, we can use numerical simulations to assess the algorithm's performance. We study {\parity} RQAOA's performance numerically for ${p=1}$ layer, where analytical formulas for the objective function are available. Following Ref.~\cite{Wybo_Leib_Quantum2024}, we consider different instances of the signed Max-Cut problem on a complete graph with random weights ${J_{jk} \in \{-1, 1\}}$ (see Sec.~\ref{subsec:SK-model}).
In Fig.~\ref{fig:rqaoa}, we compare single-layer QAOA and RQAOA approximation ratios resulting from a {\parity} implementation with those resulting from a {\vanilla} implementation, where we use $\tilde{\calC}^{\mathrm{mean}}$ as an objective function for {\parity} QAOA. As predicted in Ref.~\cite{Wybo_Leib_Quantum2024}, {\parity} QAOA produces approximation ratios below  {\vanilla} QAOA. However, using RQAOA results in higher approximation ratios for both implementations. In particular, {\parity} RQAOA performs as well as {\vanilla} RQAOA, despite the latter requiring deeper circuits. 

\section{Conclusion and Outlook}\label{sec:conclusion}

In this work, we studied the performance and characteristics of {\parity} QAOA and compared them with those of {\vanilla} QAOA. The analysis focused on the signed Max-Cut problem for the Sherrington-Kirkpatrick spin glass and random regular graph instances, leveraging Clifford circuits to estimate {\parity} QAOA's performance on large systems.

Our QAOA numerical simulations highlight how the choice of the decoding strategy affects {\parity} QAOA's performance. In particular, we found that selecting an optimal spanning tree during each measurement decoding significantly enhances the algorithm's performance. Taking this into account, we verified that {\parity} QAOA outperforms {\vanilla} QAOA by achieving higher approximation ratios with a smaller circuit depth on problems with up to twenty-one variables. 

For larger systems, the local nature of interactions in the {\parity} architecture is expected to hinder the algorithm's performance by limiting the spread of information across the system. Thus, following Ref.~\cite{Bravyi_Tang_PRL2020}, we introduced {\parity} RQAOA to circumvent this limitation, and we benchmarked the algorithm's single-layer implementation. We found that despite only requiring constant-depth quantum circuits, {\parity} RQAOA achieves the same approximation ratio as recursive {\vanilla} QAOA, where the circuit depth scales linearly with the problem size.

Our findings suggest that {\parity} QAOA offers a promising alternative to {\vanilla} QAOA for solving large and complex optimization problems. As it has different demands on quantum hardware, it is essential to consider the specific characteristics of the target platform (2-qubit gate fidelity, coherence times, idle fidelity...) when selecting the appropriate implementation.
In addition, the algorithm's performance strongly depends on the details of the implementation and the classical post-processing methods considered (e.g., decoding and recursive strategies). Therefore, future research should focus on further developing these techniques and extending them to a broader range of problems, including graphs with non-discrete weights. In these studies, classical simulations of Clifford circuits may offer resource-efficient strategies to compare different QAOA variants and architectures~\cite{MunosArias_Blais_PRR2024, Herrman_ScientificReports2022, wilkie_Herrman_arXiv2024}. 
Moreover, the Clifford circuit initialization method proposed in Ref.~\cite{Cheng_Kim_arXiv2022} may provide low-energy starting points for QAOA, ultimately improving the algorithm's efficiency.

Finally, we note that the analysis presented does not account for quantum noise, which still plays a crucial role in current implementations of quantum algorithms. However, as reported in Ref.~\cite{Weidinger2023}, we expect the redundant encoding of information in the {\parity} architecture and the spanning tree decoding to mitigate quantum noise, potentially making QAOA less susceptible to gate errors and decoherence effects. A detailed investigation to verify these statements and test the algorithm's scalability on physical hardware remains an open area for further research.

\vspace{5mm}
A Python code implementation of the Parity QAOA Clifford circuits considered in this work is available online \cite{CliffordCircuit_GitlabTutorial}.

\section*{Acknowledgements}
We thank Philipp Aumann, Roeland ter Hoeven, Barry Mant, Anette Messinger, and Leo Stenzel for useful discussions. This study was supported by NextGenerationEU via FFG and Quantum Austria (FFG Project No. FO999896208).
This research was funded in part by the Austrian Science Fund (FWF) under Grant-DOI 10.55776/F71 and Grant-DOI 10.55776/Y1067.
This project was funded within the QuantERA II Programme that has received funding from the European Union’s Horizon 2020 research and innovation programme under Grant Agreement No. 101017733. 
This publication has received funding under Horizon Europe programme HORIZON-CL4-2022-QUANTUM-02-SGA via the project 101113690 (PASQuanS2.1)
This study was supported by the Federal Ministry for Economic Affairs and Climate Action through project QuaST.
For the purpose of open access, the author has applied a CC BY public copyright license to any Author Accepted Manuscript version arising from this submission.

\appendix

\setlength{\tabcolsep}{4pt}
\renewcommand{\arraystretch}{1.5}
\begin{table}
\centering
\begin{tabular}{c | c || c | c} 
  \multicolumn{2}{c ||}{even $p$} &   \multicolumn{2}{c}{odd $p$} \\ [0.5ex] 
  \hline
  $l$ & $\gamma_{l}, \Omega_{l
  }, \beta_{l}$ & $l$ & $\gamma_{l}, \Omega_{l}, \beta_{l}$ \\ [0.5ex] 
  \hline\hline
  $1,\dots\frac{p-2}{2}$ & $0, \frac{1}{4}, \frac{1}{4}$ & $1,\dots\frac{p-1}{2} $ & $0, \frac{1}{4}, \frac{1}{4}$ \\
  \hline
  $\frac{p}{2}$ & $0, \frac{1}{4}, \frac{1}{2}$ & $\frac{p+1}{2}$ & $ 0/\frac{1}{2}, \frac{1}{2}, \frac{1}{4}$ \\
  \hline
    $\frac{p+2}{2},\dots(p-1)$ & $0/\frac{1}{2}, \pm\frac{1}{4}, \frac{1}{4}$ & $\frac{p+3}{2},\dots(p-1)$ & $ 0/\frac{1}{2}, \pm\frac{1}{4}, \frac{1}{4}$ \\
    \hline
    $p$ & $\pm\frac{1}{4}, \pm\frac{1}{4}, \frac{1}{4}$ & $p$ & $ \pm\frac{1}{4}, \pm\frac{1}{4}, \frac{1}{4}$ \\
 [1ex] 
 \hline
\end{tabular}
\caption{Classical parameter vectors for even (left) and odd (right) QAOA layers $p>1$. The table shows the index values $l=1,\dots, p$, for the parameters $(\gamma_{l}, \Omega_{l}, \beta_{l})$ (given in units of $\pi$). For $p=2$ and $p=3$ descending indexes $l$ need to be skipped.
One can obtain all classical parameter vectors at layer $p$ by combining the possible values for the $3p$ parameters.}
\label{tab:app:Parameters}
\end{table}

\section{Classical parameter vectors} \label{sec:app:classical_clifford_angles}
A classical parameter vector $(\bfgamma, \bfOmega, \bfbeta)\in \Aclassical$ applied to QAOA unitary yields a classical state 
\begin{equation}
\begin{split}
     \ket{\boldsymbol{\tilde q}}=&\ket{\tilde{q}_{1}, \tilde{q}_{2}, \dots, \tilde{q}_{K}} =\tilde{U}_x(\beta_{p})\tilde{U}_{\squaredots}(\Omega_{p})\tilde{U}_z(\gamma_{p})\times\cdots \\
    &\cdots \times\tilde{U}_x(\beta_{1})\tilde{U}_{\squaredots}(\Omega_{1})\tilde{U}_z (\gamma_{1})\ket{+}^{\otimes K},
\end{split}\label{eq:classical_state_circuit}
\end{equation}
where ${\tilde{q}_{k}\in\{0, 1\}}$ with $k=1,\dots,K$.
For {\parity} QAOA and ${p=1}$ we obtain four classical parameter vectors. Denoting those vectors with $\boldsymbol{\vartheta}=(\gamma_1, \Omega_1, \beta_1)$, they are

\begin{align}\label{eq:app:angles_for_p1_1}
 \boldsymbol{\vartheta}_1 &= \left(\frac{\pi}{4}, 0, \frac{\pi}{4}\right), \\ 
 \boldsymbol{\vartheta}_2 &= \left(\frac{\pi}{4}, \frac{\pi}{2}, \frac{\pi}{4}\right), \\
 \boldsymbol{\vartheta}_3 &= \left(-\frac{\pi}{4}, 0, \frac{\pi}{4}\right), \label{eq:app:angles_for_p1_3}\\
 \boldsymbol{\vartheta}_4 &= \left(-\frac{\pi}{4}, \frac{\pi}{2}, \frac{\pi}{4}\right). \label{eq:app:angles_for_p1_4}
\end{align}

Table~\ref{tab:app:Parameters} shows how to construct parameter vectors for ${p>1}$. Here, we distinguish whether the number of layers $p$ is even or odd. The index $l$ and the values $(\gamma_{l}, \Omega_{l}, \beta_{l})$ determine the whole set.
For $p=2$ and $p=3$ one needs to skip some rows, namely the ones where a given sequence of index $l$ is descending (first and third row for $p=2$ and third row for $p=3$).

Note that for ${l>p/2}$, there are two possible values for $\gamma_{l}$ as well as for $\Omega_{l}$ if $p$ is even, while for odd $p$ there is an extra parameter $\gamma_{l}$ at $l=(p+1)/2$. This leads to $p$ parameters which can take two values at each $p$. The combination of all possible values comprises all ${2^p}$ classical parameter vectors at each $p$. Taking into account the parameter vectors for ${q < p}$ layers (by filling up the remaining values with zeros) and evaluating the partial sum of the resulting geometric series yields
\begin{equation}
    |\Aclassical| = \sum_{q=1}^p 2^q = \frac{1-2^{p+1}}{1-2}-1 = 2^{p+1}
\end{equation}
possible classical parameter vectors for $p$ QAOA layers. 

An example for constructing the classical parameter vectors for ${p=4}$ (without accounting for the parameter vector for lower $p$) can be found in Tab.~\ref{tab:app:Parameters_p4}.

\setlength{\tabcolsep}{2.9pt}
\begin{table}
\centering
\begin{tabular}{c c c | c c c | c c c | c c c}
 $\gamma_{1}$ & $\Omega_{1}$ & $\beta_{1}$ & $\gamma_{2}$ & $\Omega_{2}$ & $\beta_{2}$ & \cellcolor{gray!15}$\gamma_{3}$ & \cellcolor{gray!15}$\Omega_{3}$ & $\beta_{3}$ &  \cellcolor{gray!15}$\gamma_{4}$ & \cellcolor{gray!15}$\Omega_{4}$ & $\beta_{4}$\\ [0.5ex] 
 \hline\hline 
   $0$ & $\frac{1}{4}$ & $\frac{1}{4}$ &  $0$ & $\frac{1}{4}$ & $ \frac{1}{2}$ & $0$ & $\frac{1}{4}$ & $\frac{1}{4}$ & $\frac{1}{4}$ & $\frac{1}{4}$ & $\frac{1}{4}$\\ 
   $0$ & $\frac{1}{4}$ & $\frac{1}{4}$ &  $0$ & $\frac{1}{4}$ & $ \frac{1}{2}$ & $0$ & $\frac{1}{4}$ & $\frac{1}{4}$ & $\frac{1}{4}$ & \cellcolor{gray!15}$-\frac{1}{4}$ & $\frac{1}{4}$\\ 
  $0$ & $\frac{1}{4}$ & $\frac{1}{4}$ &  $0$ & $\frac{1}{4}$ & $ \frac{1}{2}$ & $0$ & $\frac{1}{4}$ & $\frac{1}{4}$ & \cellcolor{gray!15}$-\frac{1}{4}$ & $\frac{1}{4}$ & $\frac{1}{4}$\\ 
  $0$ & $\frac{1}{4}$ & $\frac{1}{4}$ &  $0$ & $\frac{1}{4}$ & $ \frac{1}{2}$ & $0$ & $\frac{1}{4}$ & $\frac{1}{4}$ &\cellcolor{gray!15} $-\frac{1}{4}$ & \cellcolor{gray!15}$-\frac{1}{4}$ & $\frac{1}{4}$\\ 
  \hline
  $0$ & $\frac{1}{4}$ & $\frac{1}{4}$ &  $0$ & $\frac{1}{4}$ & $ \frac{1}{2}$ & $0$ &\cellcolor{gray!15} $-\frac{1}{4}$ & $\frac{1}{4}$ & $\frac{1}{4}$ & $\frac{1}{4}$ & $\frac{1}{4}$\\ 
    $0$ & $\frac{1}{4}$ & $\frac{1}{4}$ &  $0$ & $\frac{1}{4}$ & $ \frac{1}{2}$ & $0$ & \cellcolor{gray!15}$-\frac{1}{4}$ & $\frac{1}{4}$ & $\frac{1}{4}$ & \cellcolor{gray!15}$-\frac{1}{4}$ & $\frac{1}{4}$\\ 
 $0$ & $\frac{1}{4}$ & $\frac{1}{4}$ &  $0$ & $\frac{1}{4}$ & $ \frac{1}{2}$ & $0$ & \cellcolor{gray!15}$-\frac{1}{4}$ & $\frac{1}{4}$ & \cellcolor{gray!15}$-\frac{1}{4}$ & $\frac{1}{4}$ & $\frac{1}{4}$\\ 
  $0$ & $\frac{1}{4}$ & $\frac{1}{4}$ &  $0$ & $\frac{1}{4}$ & $ \frac{1}{2}$ & $0$ & \cellcolor{gray!15}$-\frac{1}{4}$ & $\frac{1}{4}$ & \cellcolor{gray!15}$-\frac{1}{4}$ & \cellcolor{gray!15}$-\frac{1}{4}$ & $\frac{1}{4}$\\ 
  \hline
  $0$ & $\frac{1}{4}$ & $\frac{1}{4}$ &  $0$ & $\frac{1}{4}$ & $ \frac{1}{2}$ & \cellcolor{gray!15}$\frac{1}{2}$ & $\frac{1}{4}$ & $\frac{1}{4}$ & $\frac{1}{4}$ & $\frac{1}{4}$ & $\frac{1}{4}$\\ 
    $0$ & $\frac{1}{4}$ & $\frac{1}{4}$ &  $0$ & $\frac{1}{4}$ & $ \frac{1}{2}$ & \cellcolor{gray!15}$\frac{1}{2}$ & $\frac{1}{4}$ & $\frac{1}{4}$ & $\frac{1}{4}$ & \cellcolor{gray!15}$-\frac{1}{4}$ & $\frac{1}{4}$\\ 
  $0$ & $\frac{1}{4}$ & $\frac{1}{4}$ &  $0$ & $\frac{1}{4}$ & $ \frac{1}{2}$ & \cellcolor{gray!15}$\frac{1}{2}$ & $\frac{1}{4}$ & $\frac{1}{4}$ & \cellcolor{gray!15}$-\frac{1}{4}$ & $\frac{1}{4}$ & $\frac{1}{4}$\\ 
$0$ & $\frac{1}{4}$ & $\frac{1}{4}$ &  $0$ & $\frac{1}{4}$ & $ \frac{1}{2}$ & \cellcolor{gray!15}$\frac{1}{2}$ & $\frac{1}{4}$ & $\frac{1}{4}$ & \cellcolor{gray!15}$-\frac{1}{4}$ & \cellcolor{gray!15}$-\frac{1}{4}$ & $\frac{1}{4}$\\ 
  \hline
   $0$ & $\frac{1}{4}$ & $\frac{1}{4}$ &  $0$ & $\frac{1}{4}$ & $ \frac{1}{2}$ & \cellcolor{gray!15}$\frac{1}{2}$ & \cellcolor{gray!15}$-\frac{1}{4}$ & $\frac{1}{4}$ & $\frac{1}{4}$ & $\frac{1}{4}$ & $\frac{1}{4}$\\ 
   $0$ & $\frac{1}{4}$ & $\frac{1}{4}$ &  $0$ & $\frac{1}{4}$ & $ \frac{1}{2}$ & \cellcolor{gray!15}$\frac{1}{2}$ & \cellcolor{gray!15}$-\frac{1}{4}$ & $\frac{1}{4}$ & $\frac{1}{4}$ & \cellcolor{gray!15}$-\frac{1}{4}$ & $\frac{1}{4}$\\ 
  $0$ & $\frac{1}{4}$ & $\frac{1}{4}$ &  $0$ & $\frac{1}{4}$ & $ \frac{1}{2}$ & \cellcolor{gray!15}$\frac{1}{2}$ & \cellcolor{gray!15}$-\frac{1}{4}$ & $\frac{1}{4}$ & \cellcolor{gray!15}$-\frac{1}{4}$ & $\frac{1}{4}$ & $\frac{1}{4}$\\ 
   $0$ & $\frac{1}{4}$ & $\frac{1}{4}$ &  $0$ & $\frac{1}{4}$ & $ \frac{1}{2}$ & \cellcolor{gray!15}$\frac{1}{2}$ & \cellcolor{gray!15}$-\frac{1}{4}$ & $\frac{1}{4}$ & \cellcolor{gray!15}$-\frac{1}{4}$ & \cellcolor{gray!15}$-\frac{1}{4}$ & $\frac{1}{4}$\\ 
 [1ex] 
 \hline
\end{tabular}
\caption{Classical parameter vectors for $p=4$ layers (excluding the ones from $p<4$). Values are given in units of $\pi$. The vectors are constructed by the method given in Tab.~\ref{tab:app:Parameters}. This table suggests two distinct values for the 4 parameters $\gamma_{3}$, $\Omega_{3}$, $\gamma_{4}$ and $\Omega_{4}$ (marked in gray in the header). 
All possible combinations of the four values lead to a total of ${2^4=16}$ vectors. A change in the background color of a value (e.g., white to gray and vice versa) in the table indicates that the value is different than the one in the previous row. If there is no gray marked cell in a column, the values are constant for all parameter sets.}
\label{tab:app:Parameters_p4}
\end{table}

\begin{figure}
    \centering
    \includegraphics[width=0.8\linewidth]{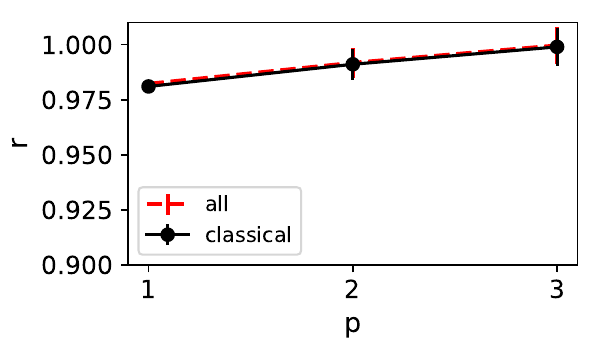}
    \caption{Approximation ratio achieved by using all possible Clifford angles (red dashed line) compared to using Classical Clifford parameters only (black solid line). Results shown are the average over $600$ random generated instances of size $N=7$. Errorbars indicate twice the standard error of the mean.}
    \label{fig:app:AllCLifford_vs_ClassicalOnly}
\end{figure}

Figure~\ref{fig:app:AllCLifford_vs_ClassicalOnly} shows that there is no significant difference in the approximation ratio $r$ if all possible Clifford parameter vectors or if only Classical Clifford parameter vectors are used. This is due to the fact that the majority of Clifford parameter vectors return an equal superposition of all possible states. Neglecting those Clifford parameter vectors not only reduces the number of considered parameter vectors for the lower bound (both still scale exponentially in $p$) but also reduces the number of measurements needed. Classical Clifford vectors generate a circuit that returns a classical state, which requires a single measurement only, whereas a superposition state requires at least 1.000 measurements to estimate the energy.

\section{Classical states}\label{sec:app:classical_clifford_states}
In this section, we focus on the classical states that are obtained from the circuits generated by the classical parameter vectors.

For ${p=1}$, the four classical states $\boldsymbol{\tilde q}_1$, $\boldsymbol{\tilde q}_2$, $\boldsymbol{\tilde q}_3$ and $\boldsymbol{\tilde q}_4$ correspond to the parameter vectors from Eqs.~\eqref{eq:app:angles_for_p1_1}-\eqref{eq:app:angles_for_p1_4}.
The state $\boldsymbol{\tilde q}_1$ is the ground state of the negative local-field Hamiltonian $-\tilde{H}_P$, Eq.~\eqref{eq:local_field_hamiltonian}. All qubits $\tilde{\sigma}_{\nu}^z$ are aligned to the fields $-\tilde{J}_\nu$. 
In contrast, the state $\boldsymbol{\tilde q}_3$ has all qubits $\tilde{\sigma}_{\nu}^z$ aligned to the fields $+\tilde{J}_\nu$, indicating that $\boldsymbol{\tilde q}_3$ is the ground state of the local-field Hamiltonian $\tilde{H}_P$. The states $\boldsymbol{\tilde q}_1$ and $\boldsymbol{\tilde q}_3$ differ by a global spin flip.
Comparing the parameter vectors $\boldsymbol{\vartheta}_1$ and $\boldsymbol{\vartheta}_3$ shows that they differ by a sign in $\gamma_1^\prime$. 
In addition, both constraint values $\Omega_1^\prime$ are zero, which implies that both circuits perform only local rotations. 
Continuing to the sets $\boldsymbol{\vartheta}_2$ and $\boldsymbol{\vartheta}_4$, which differ in the sign of $\gamma_1$, their corresponding states $\boldsymbol{\tilde q}_2$ and $\boldsymbol{\tilde q}_4$ show the same behavior as the previous states and differ by a global spin flip. 
In more detail, for the state $\boldsymbol{\tilde q}_2$, qubits that are in an uneven amount of constraints are aligned, while the rest is anti-aligned to the fields $\tilde{J}_{\nu}$. Table~\ref{tab:app:Parameters_p1_examples} shows the configuration of the classical states for the signed Max-Cut problem on a complete graph and different system sizes $N$. Red (gray) circles denote qubits $\tilde{\sigma}_{\nu}^z$ that are anti-aligned (aligned) to the local field $\tilde{J}_{\nu}$.

Every classical state can be specified by the orientation of the qubits $\tilde{\sigma}_{\nu}^z$ to their field $\tilde{J}_{\nu}$. The pattern of aligned and anti-aligned qubits for a specific size, for example ${N=4}$, can be extended to larger system sizes ${N>4}$.
Table~\ref{tab:app:Parameters_p2_examples} displays the states occurring for ${p=2}$, which shows that the state patterns become more complex for ${p>1}$.

\setlength{\tabcolsep}{1pt}
\begin{table}
\centering
\begin{tabular}{c | c | c | c | c} 
 N & $\frac{1}{4}, 0, \frac{1}{4}$ & $\frac{1}{4}, \frac{1}{2}, \frac{1}{4}$ & $-\frac{1}{4}, 0, \frac{1}{4}$ & $-\frac{1}{4}, \frac{1}{2}, \frac{1}{4}$\\ [0.5ex] 
 \hline \hline
  \parbox[m]{.6cm}{4}  & \parbox[m]{1.8cm}{\includegraphics[width=50pt]{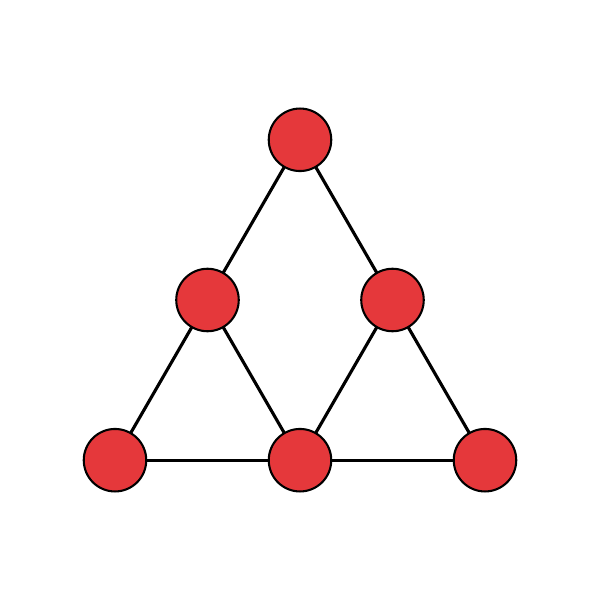}} & \parbox[m]{1.85cm}{\includegraphics[width=50pt]{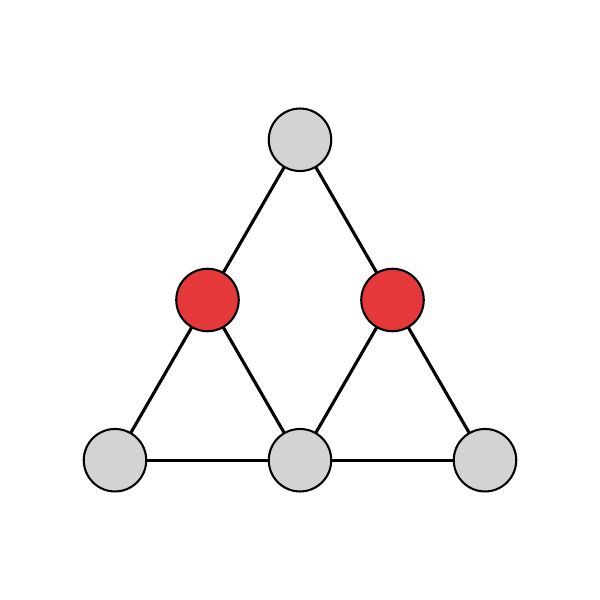}} &  \parbox[m]{1.85cm}{\includegraphics[width=50pt]{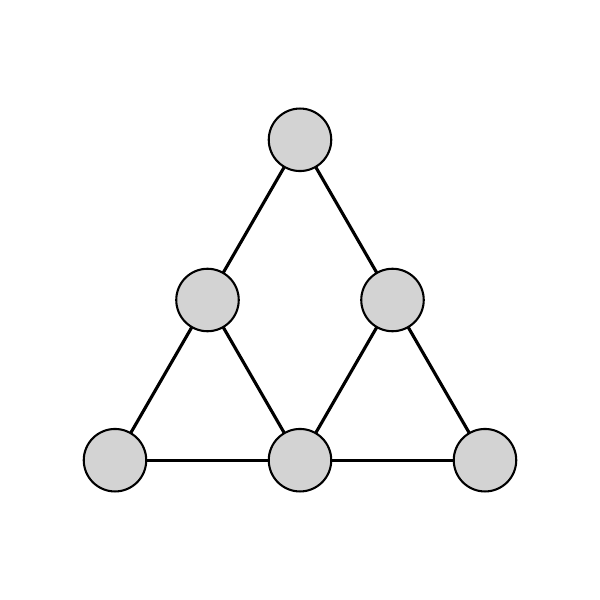}}& \parbox[m]{1.85cm}{\includegraphics[width=50pt]{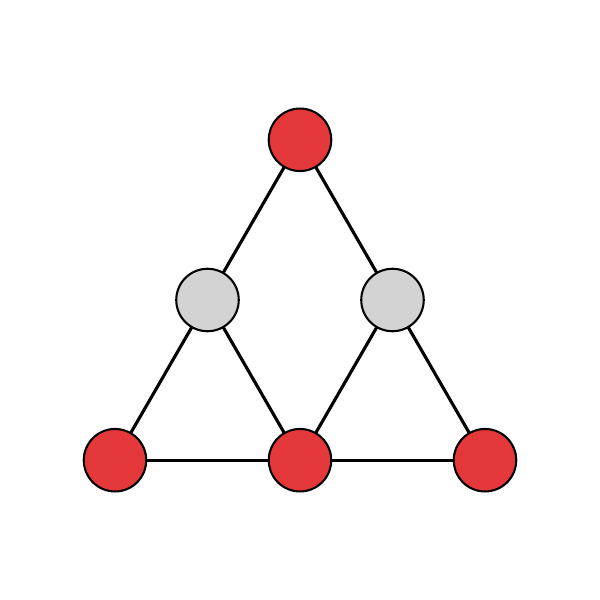}} \\
  \hline
 \parbox[m]{.6cm}{5} & \parbox[m]{1.8cm}{\includegraphics[width=50pt]{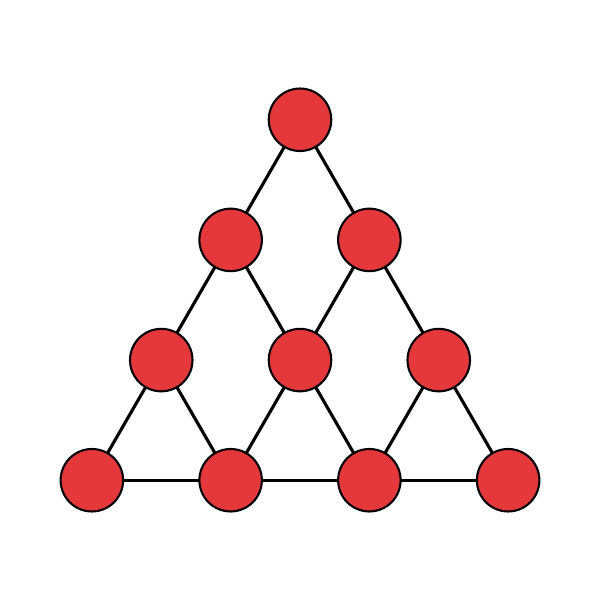}} & \parbox[m]{1.8cm}{\includegraphics[width=50pt]{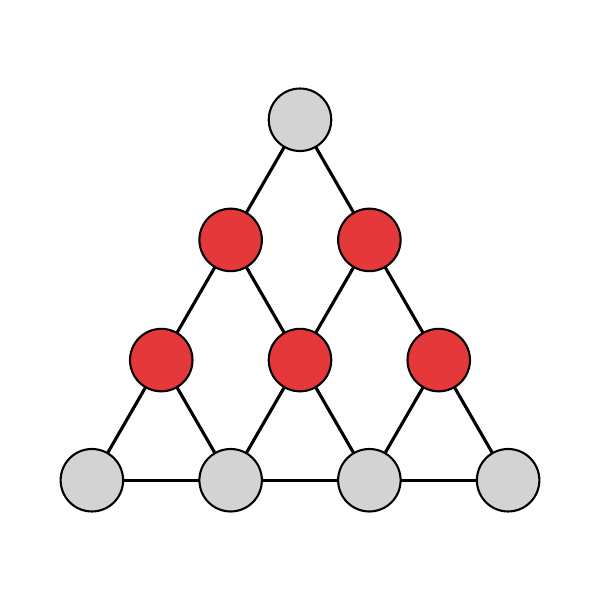}} &  \parbox[m]{1.8cm}{\includegraphics[width=50pt]{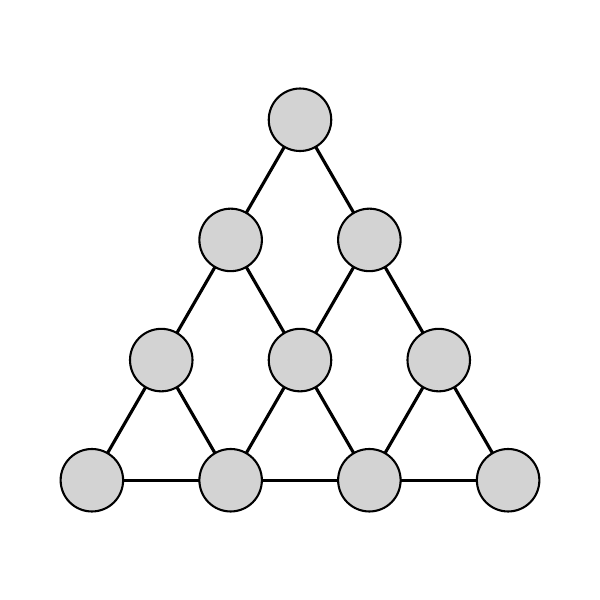}} & \parbox[m]{1.8cm}{\includegraphics[width=50pt]{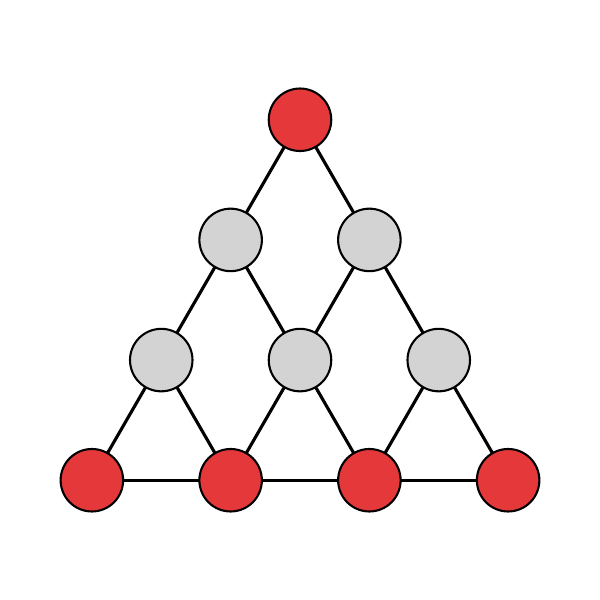}} \\ 
 \hline
\parbox[m]{.6cm}{6} & \parbox[m]{1.8cm}{\includegraphics[width=50pt]{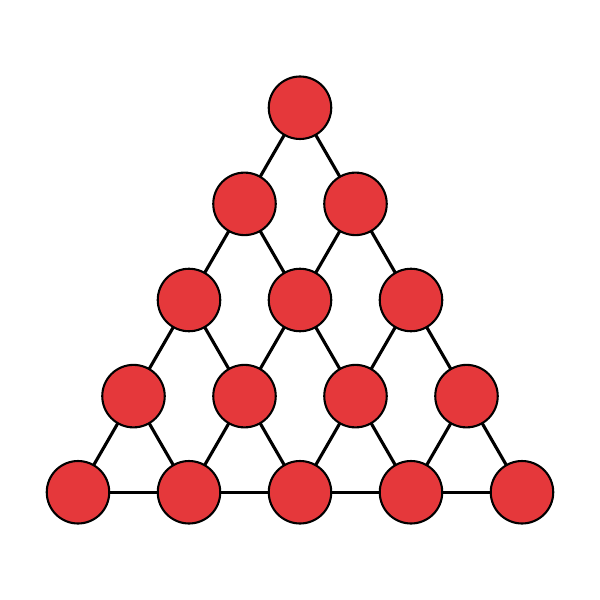}} & \parbox[m]{1.8cm}{\includegraphics[width=50pt]{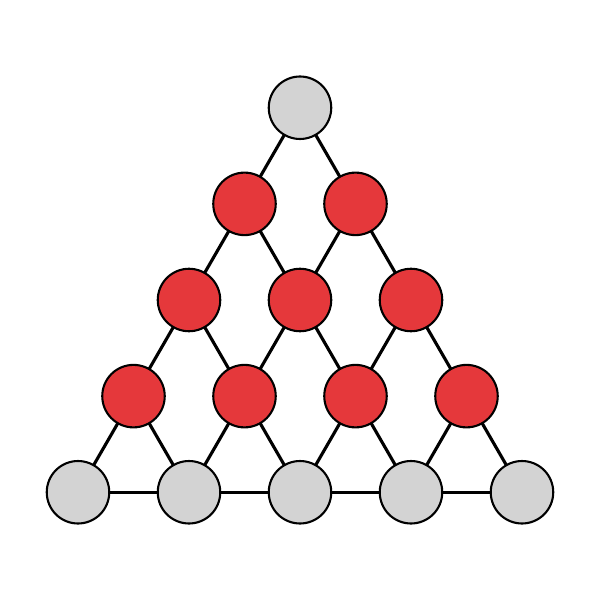}} &  \parbox[m]{1.8cm}{\includegraphics[width=50pt]{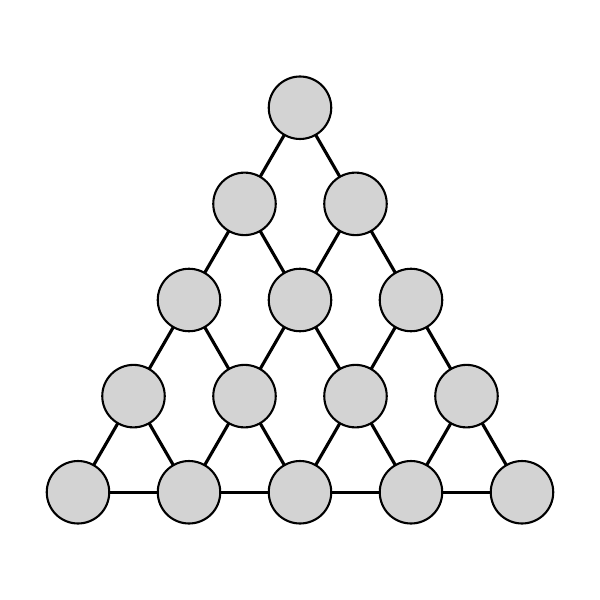}} & \parbox[m]{1.8cm}{\includegraphics[width=50pt]{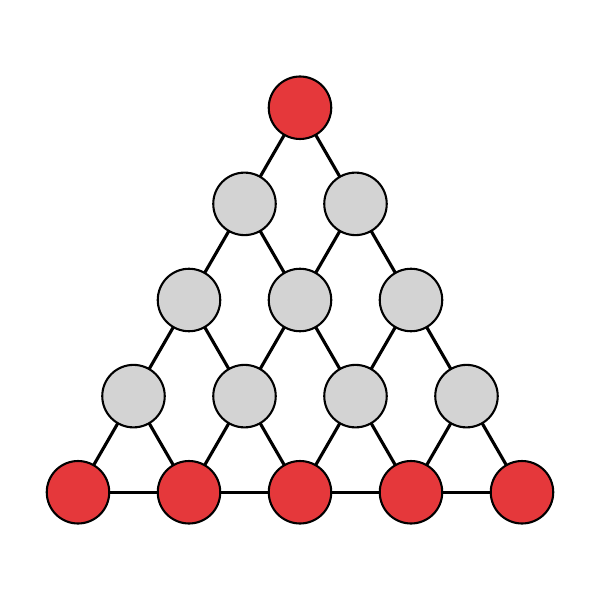}} \\ [1ex] 
 \hline
\end{tabular}
\caption{Classical parameter vectors in units of $\pi$ for ${p=1}$ with corresponding classical states for different logical qubit numbers $N$ for a complete graph. Red (gray) circles denote qubits $\tilde{\sigma}_{\nu}^z$ that are anti-aligned (aligned) to their local field $\tilde{J}_{\nu}$. The first parameter set $(\frac{1}{4}, 0, \frac{1}{4})$ leads to the inverted local-field ground state, while the third set, $(-\frac{1}{4}, 0, \frac{1}{4})$ yields the local-field ground state. Also, the second and fourth sets differ by a global spin flip.}
\label{tab:app:Parameters_p1_examples}
\end{table}

\begin{table}
\centering
\begin{tabular}{c | c | c | c | c} 
 N & $\vartheta_{f}, \frac{1}{4}, \frac{1}{4}, \frac{1}{4}$ & $\vartheta_{f}, \frac{1}{4}, -\frac{1}{4}, \frac{1}{4}$ & $\vartheta_{f}, -\frac{1}{4}, \frac{1}{4}, \frac{1}{4}$ & $\vartheta_{f}, -\frac{1}{4}, -\frac{1}{4}, \frac{1}{4}$\\ [0.5ex] 
 \hline \hline
  \parbox[m]{.6cm}{4}  & \parbox[m]{1.8cm}{\includegraphics[width=50pt]{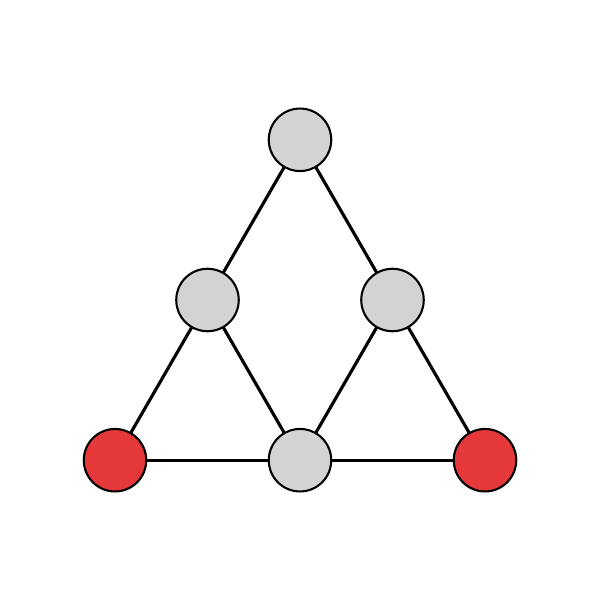}} & \parbox[m]{1.8cm}{\includegraphics[width=50pt]{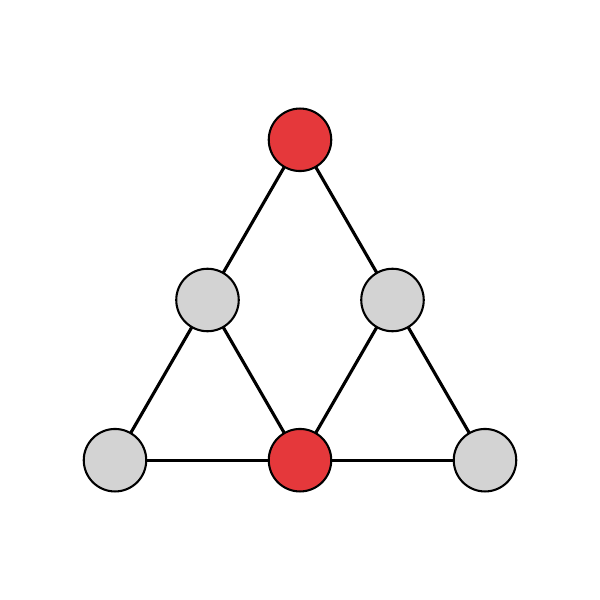}} &  \parbox[m]{1.8cm}{\includegraphics[width=50pt]{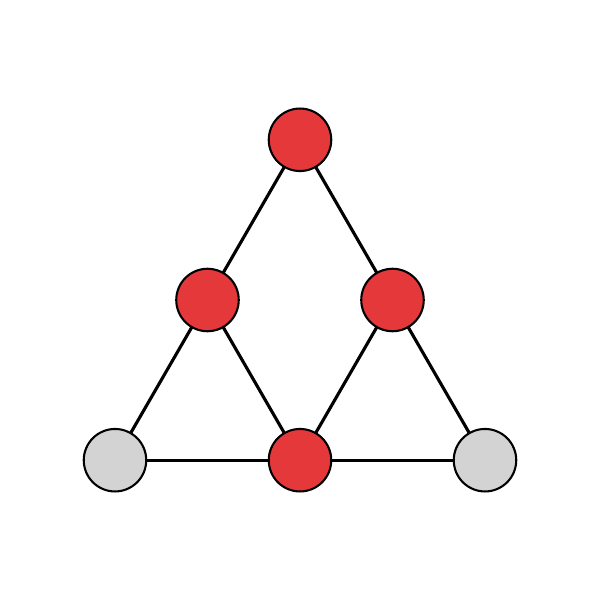}}& \parbox[m]{1.8cm}{\includegraphics[width=50pt]{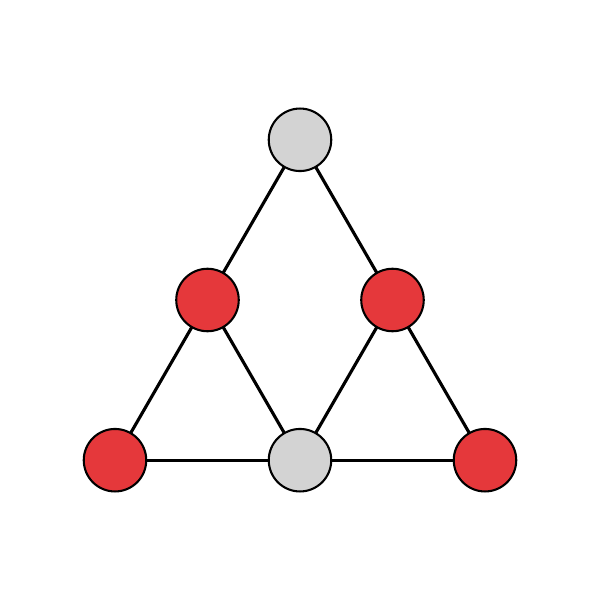}} \\
  \hline
 \parbox[m]{.6cm}{5}  & \parbox[m]{1.8cm}{\includegraphics[width=50pt]{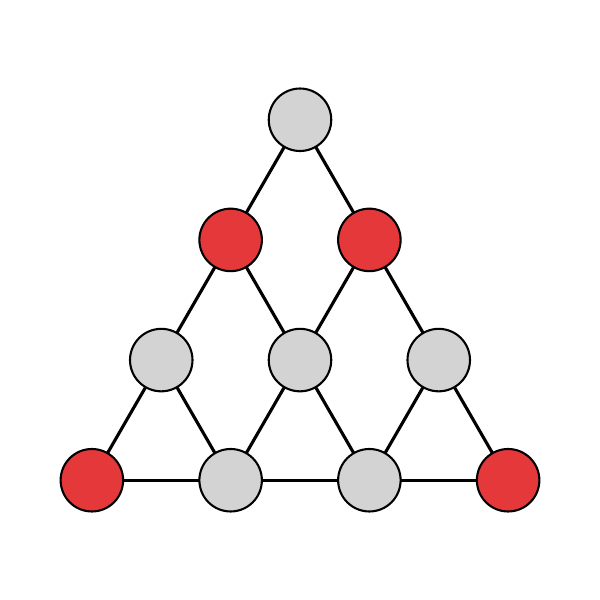}} & \parbox[m]{1.8cm}{\includegraphics[width=50pt]{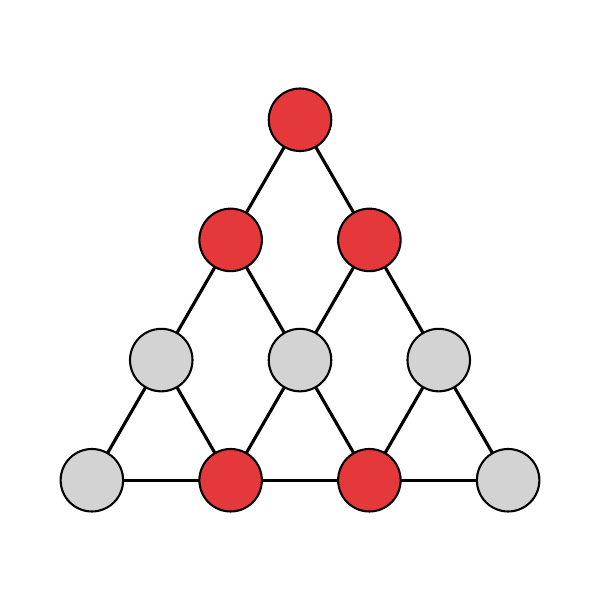}} &  \parbox[m]{1.8cm}{\includegraphics[width=50pt]{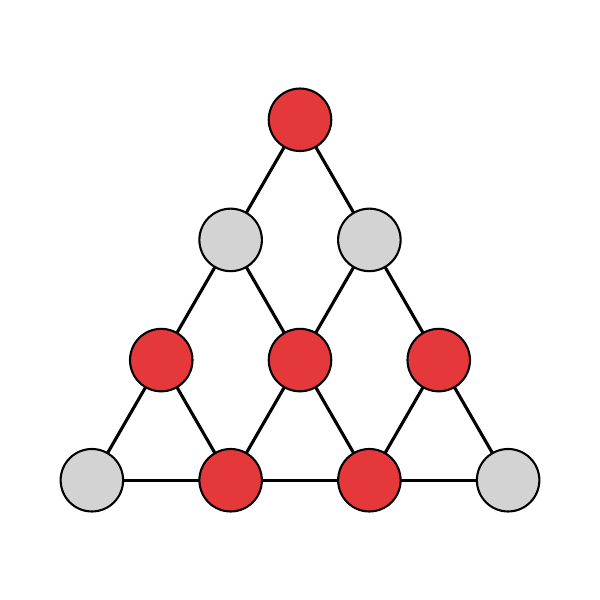}} & \parbox[m]{1.8cm}{\includegraphics[width=50pt]{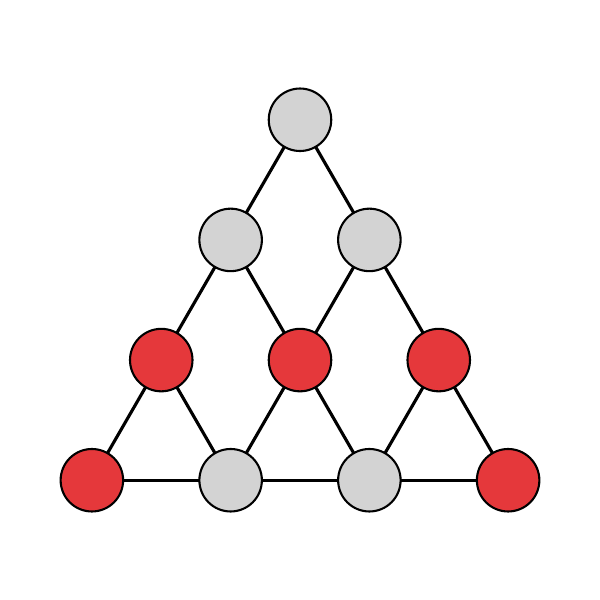}} \\ 
 \hline
\parbox[m]{.6cm}{6}  & \parbox[m]{1.8cm}{\includegraphics[width=50pt]{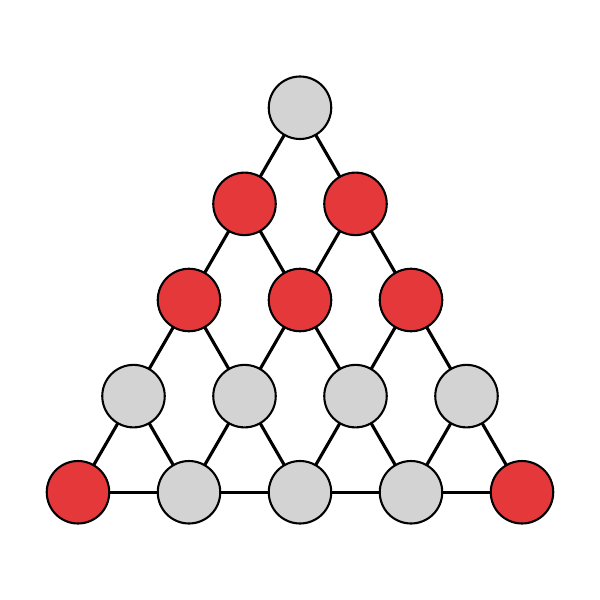}} & \parbox[m]{1.8cm}{\includegraphics[width=50pt]{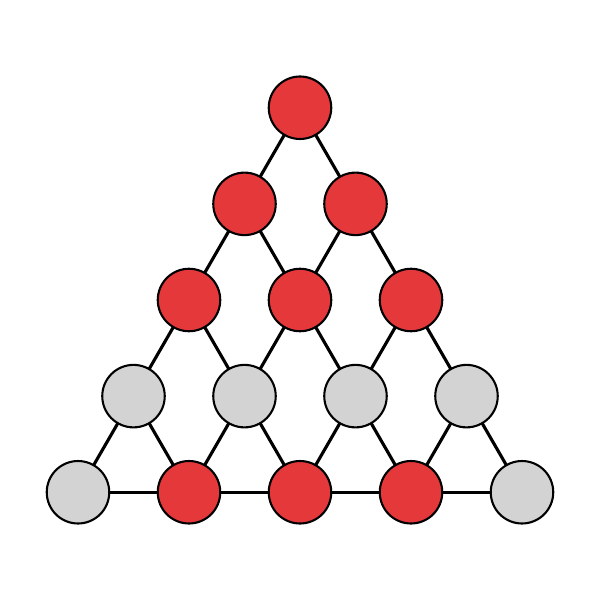}} &  \parbox[m]{1.8cm}{\includegraphics[width=50pt]{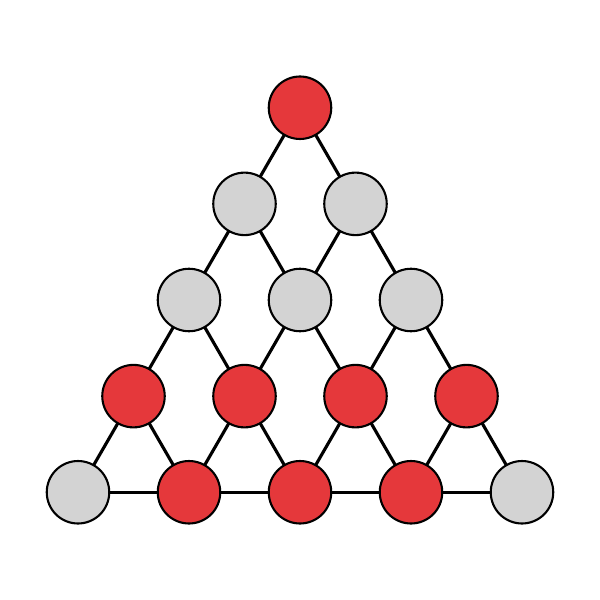}} & \parbox[m]{1.8cm}{\includegraphics[width=50pt]{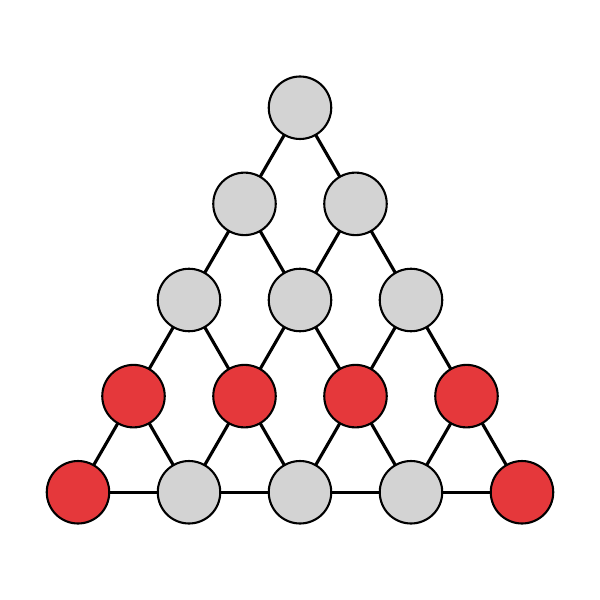}} \\ [1ex] 
 \hline
\end{tabular}
\caption{Classical states obtained from classical parameter vectors for ${p=2}$ (excluding the ones already found for ${p=1}$) for a complete graph. The values are given in units of $\pi$. All parameter vectors start with a fixed set  ${\vartheta_{f}=(\gamma_{1}, \Omega_{1}, \beta_{1})=(0, \frac{1}{4}, \frac{1}{2}}$). The values for ${(\gamma_{2}, \Omega_{2}, \beta_{2})}$ are given in the header. Red (gray) circles denote qubits $\tilde{\sigma}_{\nu}^z$ that are anti-aligned (aligned) to the their field $\tilde{J}_{\nu}$.
Negating $\gamma_p$ leads to an inverted outcome. This is the same behavior as for ${p=1}$.}
\label{tab:app:Parameters_p2_examples}
\end{table}

\section{Decoding classical states}\label{sec:app:decoding_classical_states}
\begin{figure}
    \centering
    \includegraphics[width=\columnwidth]{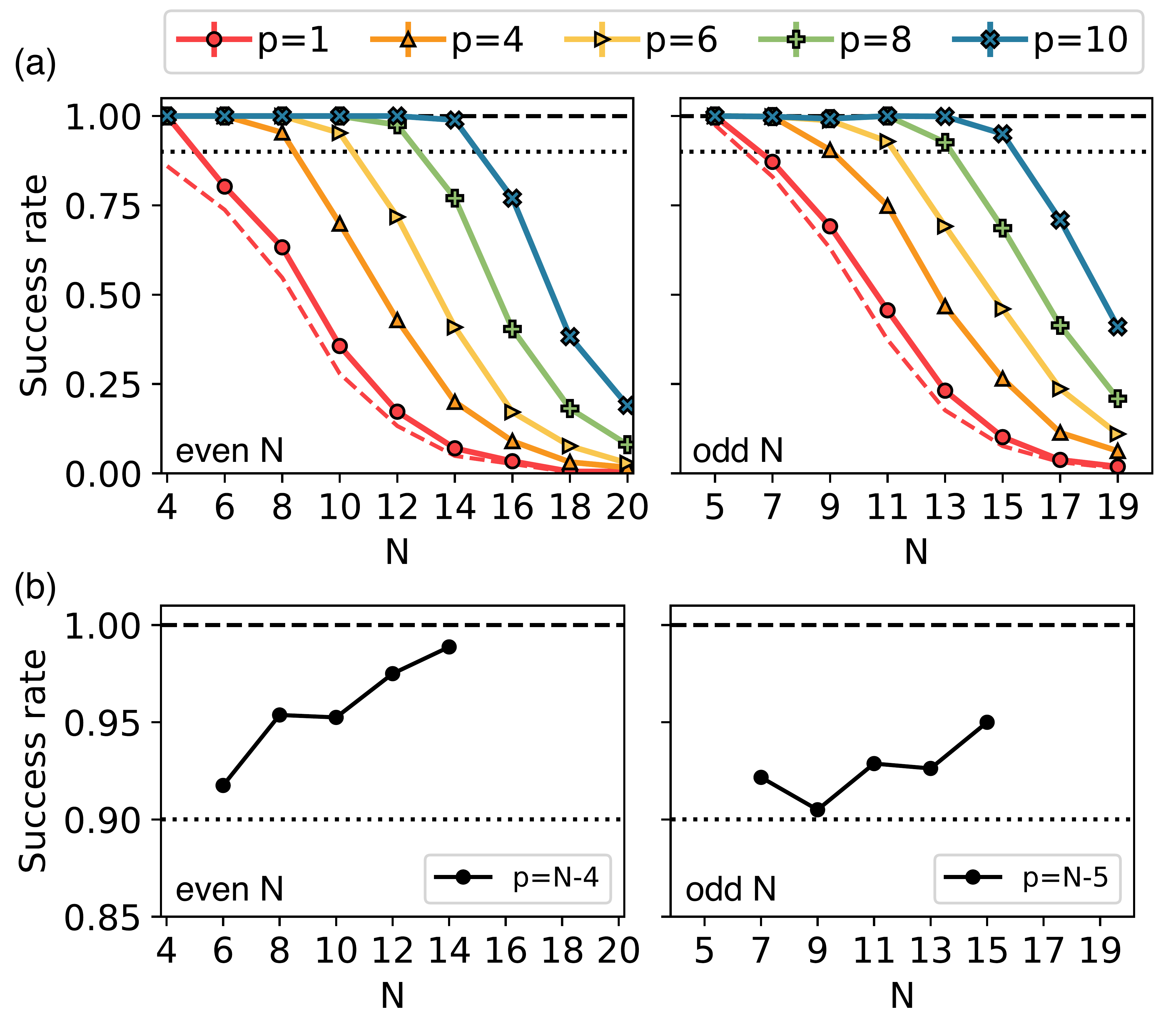}
    \caption{(a) Success rate of decoding the classical states of randomly generated problem instances for even (left) and odd (right) qubit numbers $N$ with $M=N$ for different values of $p$ for the signed Max-Cut problem on a complete graph. The red dashed line indicates the success rate of the ground state of $\tilde{H}_P$, Eq.~\eqref{eq:local_field_hamiltonian} (or $\boldsymbol{\tilde q}_3$). 
    The horizontal black dotted line marks the success rate of $0.9$. (b) Decoding success rate achieved when increasing $p$ linearly with $N$ (see legend in the subfigures). 
    Results in each figures represent the averages over 50/200/400/600/800 simulated instances for N=4/5/6/7/$>$7. The error bars, representing twice the standard error of the mean, are not visible due to their small size.}
    \label{fig:app:Succees_rate_vs_N}
\end{figure}

In this section, we address the question of how often decoding a classical state, obtained by the classical parameter vectors, yields the ground state for the signed Max-Cut problem on a complete graph. As outlined in the main text, we use the logical lines $T^{(l)}$ as spanning trees for decoding. 
We generate several problem instances with ${J_{ij}\in\{-1, 1\}}$ (exact numbers in the figure) and calculate the fraction of instances that were solved by decoding at ${p\in\{1, 2, \dots, 10\}}$.
The results (for a selection of $p$) are shown in Fig.~\ref{fig:app:Succees_rate_vs_N}(a). To avoid displaying finite-size effects, which cause higher approximation ratios for odd compared to even qubit numbers, we split the data and show even qubit numbers in the left plot, and odd qubit numbers in the right one. 
The figure shows that decoding the classical states for ${p=1}$ solves all problem instances for ${N=4}$ and ${N=5}$. That is, the success rate is equal to $1$, marked with a horizontal dashed line in the plot. However, the success rate drops with increasing $N$. By increasing $p$, the success rate increases. 
In fact, if ${p=N-2}$, all instances are solved for even qubit numbers. Interestingly, this does not hold for odd qubit numbers. There, more than $99\%$ of the instances are solved when ${p=N-2}$.
In Fig.~\ref{fig:app:Succees_rate_vs_N}(a) and (b) the dotted horizontal line marks $90\%$ success rate. The number of layers needed to obtain success rates beyond that threshold for even qubit numbers is ${p=N-4}$ and for odd qubit numbers it is ${p=N-5}$. The results for this very choice of $p$ are plotted in Figure~\ref{fig:app:Succees_rate_vs_N}(b), with even qubit numbers on the left and odd qubit numbers on the right. For even qubit numbers we observe the tendency of increasing success rate.
The investigations were made for ${p<11}$, hence, further studies should be done to check if this scaling persists for higher $p$ and $N$.

For ${N<10}$, more than half of the instances are solved by decoding the four classical states $\boldsymbol{\tilde q}_1$, $\boldsymbol{\tilde q}_2$, $\boldsymbol{\tilde q}_3$ and $\boldsymbol{\tilde q}_4$ that occur for ${p=1}$. The red dashed line (without markers) in panel (a) marks the success rate for decoding the local-field ground state $\boldsymbol{\tilde q}_3$. From the fact that the dashed line is not significantly lower than the red solid line (that combines all 4 states), we conclude that this state contributes most to the solved instances. In fact, for ${N=7}$, more than $70\%$ are solved by decoding the local-field ground state $\boldsymbol{\tilde q}_3$. Obtaining this state is trivial, the signs of the local fields $\tilde{J}_{\nu}$ are directly correlated with the spin orientation of the qubits $\sigma_\nu^z$. Indeed, the values for $\boldsymbol{\vartheta}_3$, Eq.~\eqref{eq:app:angles_for_p1_3} imply that the constraint circuit is not applied as ${\Omega_1^\prime=0}$. In Ref.~\cite{Wybo_Leib_Quantum2024}, results of {\parity} QAOA simulations with the average spanning tree energy, defined in Eq.~\eqref{eq:C_parity_mean}, as a objective function are shown. The parameters obtained there are the parameters $\boldsymbol{\vartheta}_3$.
Increasing the number of QAOA layers does not improve results as the local-field ground state $\boldsymbol{s_3}$ coincides with the ground state of $H_P$ for most instances with ${N<10}$. 

\section{Implementation and Resource Comparison Between Parity and Vanilla QAOA}
\label{sec:app:Comparison}
In this section we elaborate on the differences between vanilla and {\parity} QAOA. \\
\textit{Implementation:} Both, {\Vanilla} and {\parity} QAOA require a compilation step. While for {\Vanilla} QAOA, the problem graph is implemented directly on the hardware, necessitating a SWAP network if the graph's connectivity does not match the device topology, {\parity} QAOA requires Parity compilation~\cite{Ender2023}. For both compilations there will be a trade-off between compilation time and optimality of the compilation. For example, for {\parity} QAOA a 4-regular graph can be compiled within one step to the Parity triangle of the complete graph (Fig.~\ref{fig:Figure1}), where all parity qubits representing non existing edges in the logical graph are ancillas. More compilation time will reduce ancillas and gates needed for implementation. The same holds for {\vanilla} QAOA.\\
\textit{Qubit overhead:} For a complete graph, the qubit overhead for {\parity} QAOA is quadratic in $N$, namely $K=\frac{N(N-1)}{2}$. For other graphs, it depends on the connectivity and whether ancillary qubits are needed.
In this work, we execute {\vanilla} QAOA on K/N copies of the initial state in parallel to account for the qubit overhead in {\Parity} QAOA. After measurement of all qubits, the lowest energy obtained among all copies is used to determine the expectation value (see Appendix \ref{sec:app:simulation_details}). This improves the approximation ratio over an implementation without copies. If one does not consider copies {\vanilla} QAOA needs less qubits but more QAOA layers are needed to reach the same approximation ratio. This increases CNOT depth and gate count.\\
\textit{Post processing:} For {\Parity} QAOA we have $M$ (spanning tree) read-outs per measurement of all qubits and keep the lowest energy to evaluate the objective function. For the complete graph we considered $M=N$ spanning trees and for the 4-regular graph the number of spanning trees was $\mathcal{O}(N)$. Similarly, for {\vanilla} QAOA one obtains $K/N$ states each shot (one for each copy), where, as in {\Parity} QAOA, the state with the lowest energy is kept. The number of copies $K/N$ depends on the qubit overhead of {\Parity} QAOA which depends on the number of edges in the graph. For a complete graph the number of copies is the highest, namely $K/N=\frac{N-1}{2}$ and for the here studied 4-regular graph it is $K/N=2$. For both implementations, increasing the number of readouts/copies will improve the results, but will also increase the classical post-processing resources. \\
A comparison of the resources needed to reach an approximation ratio $r>0.95$ for both methods is shown in Tab.~\ref{tab:app:Comparison}. For the two small problems, the complete graph $N=7$ and the 4-regular graph $N=8$, the {\Parity} QAOA circuit has a depth $5$ times smaller than the {\vanilla} QAOA circuit. For the complete graph with $N=21$ we can only provide a lower bound (LB) for the approximation ratio of {\Parity} QAOA. The QAOA circuit for this lower bound exhibits a depth $2$ times smaller than the one of {\vanilla} QAOA. In terms of CNOT gate count, the {\parity} QAOA circuit for the complete graph $N=7$ comprises about $3.1$ times fewer and the circuit for the 4-regular graph $N=8$ has $2.5$ times fewer gates. However, the {\Parity} QAOA circuit for the complete graph with $N=21$ includes about $3.6$ times more CNOT gates (This is again an upper bound provided by {\Parity} LB). \\
Table~\ref{tab:app:Comparison} also provides the metric `depth times the number of qubits' which relates to the effective quantum volume proposed in Ref.~\cite{Kechedzhi_Elsevier2024}, however, in this work we did not consider noise. In this case, {\Vanilla} QAOA returns a significantly higher value.

\begin{table}
\begin{tabular}{ c | c c | c c |c c}
   &  \multicolumn{2}{c|}{SK, $N=7$} &  \multicolumn{2}{c|}{SK, $N=21$} &  \multicolumn{2}{c}{4-reg, $N=8$}  \\ 
   & P & V  & P (LB) & V  & P & V \\
   \hline
p & 1 & 2 &10 &3 & 1&3 \\  
depth & 10 & 50 & 100 & 201& 12& 60\\ 
$\frac{\text{gate  count}}{N^2}$ & 0.82 & 2.57 & 15.51 & 4.29 &0.69 & 1.78\\ 
no. of qubits & 21 & 21 &210 &210 &17 & 16 \\
no. of read-outs & 7 & 3 & 21 & 10 & 5 & 2 \\
qubits $\times$ depth & 210 & 1050 & 21000 & 42210 & 204 & 960
\end{tabular}
\caption{Resources needed for {\Vanilla} (V) and {\parity} QAOA (P) to reach an approximation ratio of $r>0.95$ for the three different problem cases shown in the main text. For {\parity} QAOA the numbers for the complete graph (SK), $N=21$ are provided by the Lower Bound (LB). The number of qubits are the (about) same for both approaches as {\vanilla} QAOA is simulated with $K/N$ copies ($\frac{K}{N}N=K$). The metric `qubits$\times$depth' relates to the effective quantum volume, however, noise was not considered in this work.}
    \label{tab:app:Comparison}
\end{table}

\section{Simulation Details}\label{sec:app:simulation_details}
In this section, we provide details on the methods and parameters we used for obtaining the simulation results presented in the main text. The description of the classical optimizer is given on the example of {\vanilla} QAOA but also holds for {\parity} QAOA.
The classical optimization of the parameters $({\bfbeta, \bfgamma})$ starts with initializing the start parameters by choosing random values from the interval $[-\frac{\pi}{2}, \frac{\pi}{2})$ (and not from $A$ as done in the Clifford pre-optimization described in Sec.~\ref{sec:clifford_circuit}). 
The parameters are updated with the Monte-Carlo technique outlined below. A new parameter set ${(\bfbeta_\text{new},\bfgamma_\text{new})}$ is generated by updating one value, e.g., ${\beta_1 \to \beta_1+ \kappa}$, where $\kappa$ is randomly drawn from a uniform distribution ${\kappa\in[-0.25, 0.25]}$. The new parameter vector is accepted if and only if  
\begin{equation}
v < \exp\left(\frac{-\calC(\bfbeta_\text{new},\bfgamma_\text{new}) + \calC(\bfbeta,\bfgamma)}{T}\right)    
\end{equation}
holds for a randomly generated value ${v\in[0, 1]}$.
The effective temperature $T$ is set to ${T=0.2}$ in this work.
After a certain amount of Monte-Carlo steps (update trials) $n_\text{mc}$, the optimization stops and we consider the lowest measured objective function $\calC(\bfbeta^*,\bfgamma^*)$ as the best outcome.
For estimating the energy expectation for a given parameter set, the quantum circuit for these parameters is executed 10.000 times. 
For every instance, $n_\text{init}$ start parameters are initialized and updated with the previously described method. The best run is considered for later statistics. The selection for $n_\text{init}$ and $n_\text{mc}$ was made so that it ensures the simulation results are close to convergence—where further increasing 
the values would not lead to significant changes in the outcomes—while also balancing the need to minimize classical simulation time.
To equalize the number of qubits needed in both QAOA approaches, we execute {\vanilla} QAOA with K/N copies in parallel. For each shot the lowest energy of the copies is used to determine the expectation value. 

The Clifford circuits are simulated using the library \emph{stim}. In order to do so, the QAOA gates need to be decomposed into the following Clifford gates:
$\tilde{U}_z(\pi/4)$ can be decomposed into $H_{xy}P_x$, where $H_{xy}$ is a variant of the Hadamard gate that swaps the X and Y axes (instead of X and Z). $P_{\{x, y, z\}}$ are the $\{x, y, z\}$-Pauli gates.
A negative sign in the argument, $\tilde{U}_z(-\pi/4)$, swaps the order of the gates to $P_xH_{xy}$.
$\tilde{U}_z(\pi/2)$ can be decomposed into $P_yP_x$. The constraint unitary $\tilde{U}_c$ is composed of CNOT chains (see Ref.~\cite{Lechner2018, Unger_Lechner_arXiv2023}) and a single $z$-Rotation that is applied in the same manner as $\tilde{U}_z$.
$\tilde{U}_x(\pi/2)$ and $\tilde{U}_x(\pi/4)$ can be decomposed into $P_zP_y$ and $P_zH_{yz}$, respectively.

\section{QAOA for non-trivial instances}\label{sec:app:Optimization}
In this section, we elaborate on the performance of {\parity} QAOA on non-trivial problem instances - instances that are not solvable by classical state decoding.

\subsection{Complete graph}\label{sec:app:optimization_SKmodel}
For a complete graph, the maximal qubit number for {\parity} QAOA that is still feasible for classical simulation with \emph{qiskit} is ${N=7}$ (${K=21}$). 
With that amount of qubits, one can generate $2^{15}$ unique problem instances. An instance is said to be unique if it differs in its $J_{ij}$ elements from others ($J_{ij}\neq J_{ij}'$) and, in addition, its local-field ground state $\boldsymbol{\tilde q}_3$ violates different parity constraints than other instances.
If $\boldsymbol{\tilde q}_3$ and ${\boldsymbol{\tilde q}_3'}$ violate the same constraints (but $J_{ij}\neq J_{ij}'$), the problems are not unique and will yield the same result when performing {\parity} QAOA under the same conditions. Therefore, the number of unique instances corresponds to the number of possible configurations of violating and fulfilling constraints. With ${N=7}$, there are $15$ constraints, leading to a total of $2^{15}$ possible constraint configurations.

\begin{figure}
    \centering
    \includegraphics[width=\columnwidth]{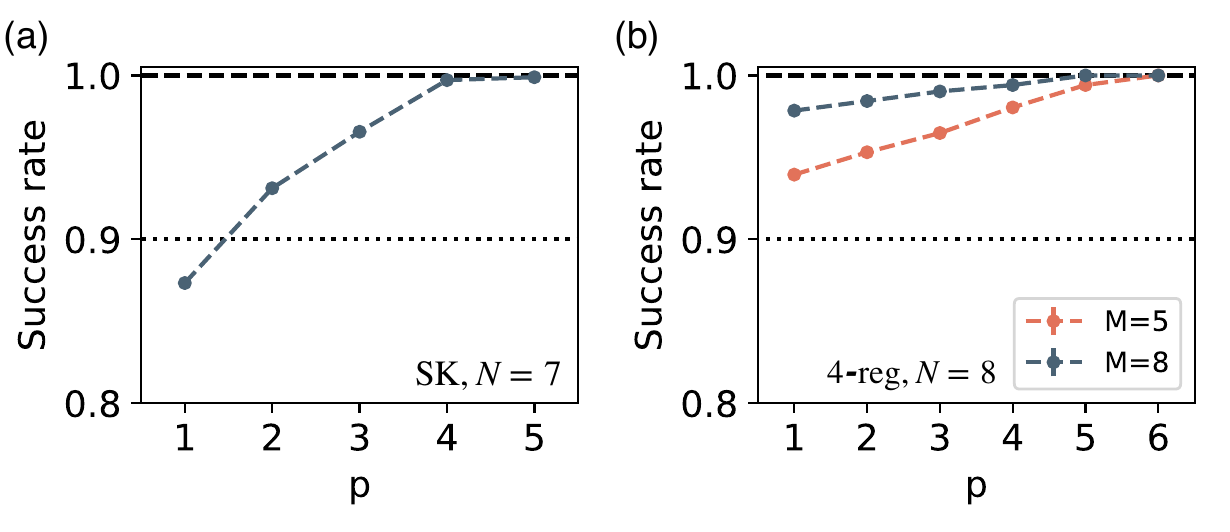}
    \caption{Success rate of decoding classical states of the signed Max-Cut problem on a (a) complete and a (b) 4-regular graph. The results shown are the average over all possible $J_{ij}$ configurations for each problem graph. Exact numbers for $p>2$ are given in Tab.~\ref{tab:app:Number_of_nontrivial_instances}.}
    \label{fig:app:SuccessRate_allJij}
\end{figure}

Data from generating and decoding all classical states with the logical lines $T^{(l)}$ for ${p=5}$ show that 36 instances are not solvable. For ${p=4}$, the number of the non-trivial instances is 94 and for ${p=3}$ it is 1200. An overview is given in Tab.~\ref{tab:app:Number_of_nontrivial_instances}. Figure~\ref{fig:app:SuccessRate_allJij}(a) shows the success rate of decoding classical states for ${p\in\{1, 2, 3, 4, 5\}}$.

\setlength{\tabcolsep}{3pt}
\begin{table}[]
    \centering
    \begin{tabular}{c|c|c|c|c}
       \multirow{2}{*}{$p$}  &   \multicolumn{2}{c|}{no. of non-solvable instances} &  \multicolumn{2}{c}{\% of total} \\\cline{2-5}
        & complete & 4-regular &  complete & 4-regular \\
       \hline
       5  &  36 & 6 (0) &  0.11 & 0.59 (0.0) \\
       4  &  94 & 20 (6) & 0.28 & 1.95 (0.59) \\
       3  &  1200 & 36 (10) &  3.66 & 3.52 (0.98)
    \end{tabular}
    \caption{Number and percentage of non-trivial instances with a given $p$ by decoding classical states for a complete graph with ${N=7}$ and ${M=7}$ and a 4-regular graph with ${N=8}$ and ${M=5}$. For the latter also values for ${M=8}$ are given in brackets. Numbers are for all possible interactions $J_{ij}$, that are  $2^{15}$ instances for the complete graph and $2^{10}$ instances for the 4-regular graph.}
    \label{tab:app:Number_of_nontrivial_instances}
\end{table}

\begin{figure}
    \centering 
    \includegraphics[width=\columnwidth] {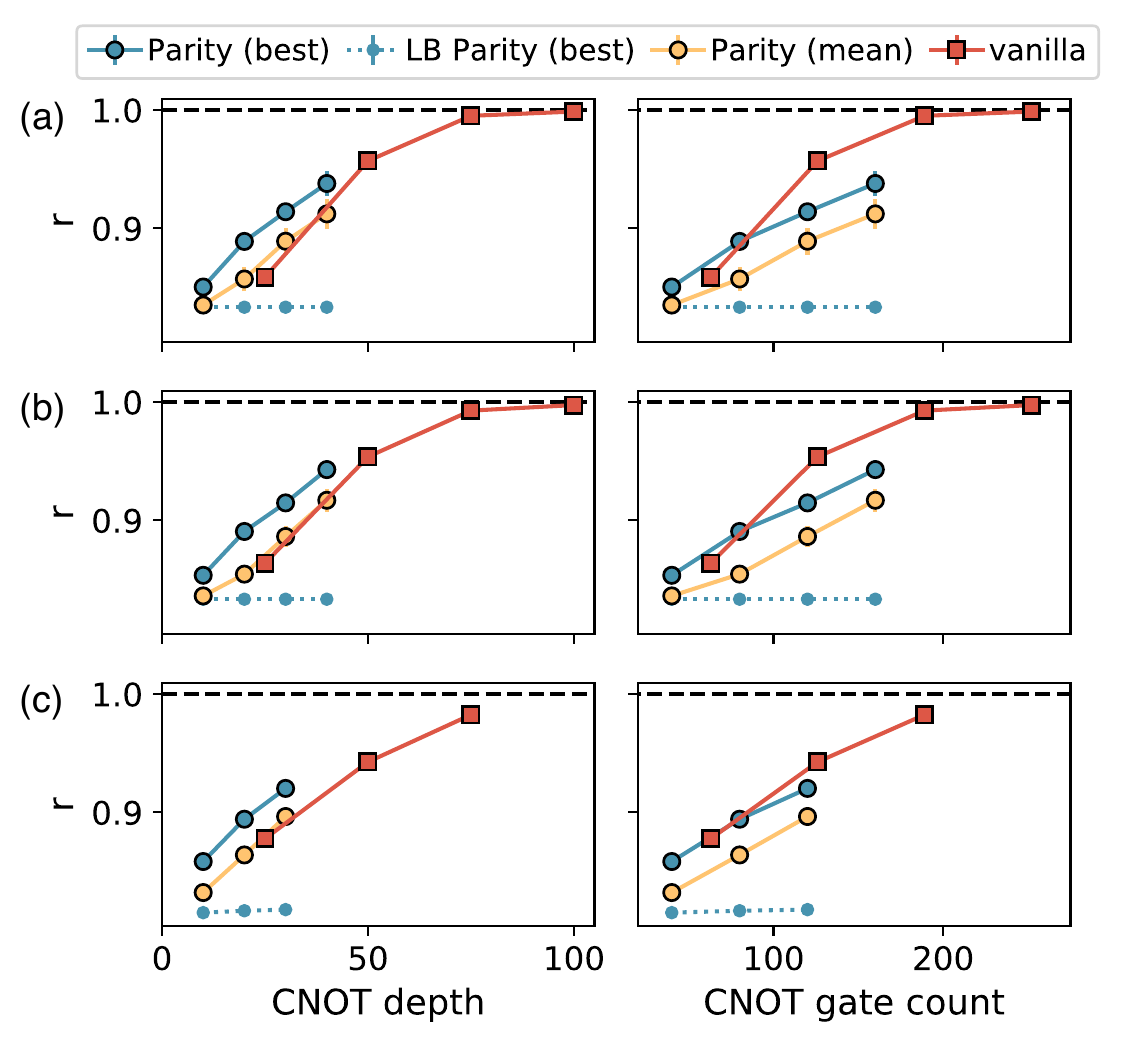}
    \caption{Approximation ratio $r$ of {\parity} and {\vanilla} QAOA for the complete graph of non-trivial instances for (a) ${p=5}$, (b) ${p=4}$ and (c) ${p=3}$ and ${N=7}$ in dependence of CNOT depth and CNOT gate count. 
    In each line, the left-most data point represents ${p=1}$, the second left-most data point represents ${p=2}$, and so forth. 
    The data points show the average of 18/29/30 non-trivial instances at ${p=5}$/$£{p=4}$/$£{p=3}$. That are all non-trivial instances (including the symmetry of the parity triangle) for each case, excluding (c) for feasibility,  whereas no instance is shown double, e.g. instances included in (a) are not considered in (b) again.
    The error bars indicate two times the standard error of the mean and are not visible due to the size of the markers. Classical optimization parameters are: ${n_\text{init}=80}$, ${n_\text{mc}=400/1000/1500/2000}$ for ${p=1/2/3/4}$.}
    \label{fig:app:Optimization_NotSolvedJij}
\end{figure}

Figure~\ref{fig:app:Optimization_NotSolvedJij} shows results of QAOA simulations of the non-trivial instances for ${p=5}$ (a), ${p=4}$ (b) and ${p=3}$~(c), whereas instances that were shown in one panel are not simulated again in another panel, e.g.,~panel (b) shows instances that are not solved at $p=4$ and are not shown in (a).
The parity representation of a complete graph is a triangle that is mirror-symmetric, which further reduces the number of unique problems. The exact numbers of unique non-trivial instances for $p$ layers can be found in Figure~\ref{tab:app:Number_of_nontrivial_instances}.
The results show that {\parity} QAOA when using $\tilde{\calC}^{\mathrm{best}}$ as the objective function, has an advantage in terms of CNOT depth but not in CNOT gate count. This is in line with the results presented in Sec.~\ref{sec:QAOA_comparison}.
The blue dotted line in each plot indicates the results obtained by decoding the classical states and represents a lower bound (LB) for the performance of {\parity} QAOA. Increasing the number of QAOA layers $p$ (and thereby increasing circuit depth and gate count) does not improve the LB of the approximation ratio $r$.
QAOA in the parity setting achieves an approximation ratio above the lower bound. With each additional layer, the approximation ratio $r$ further improves. Using the mean spanning tree energy $\tilde{\calC}^{\mathrm{mean}}$ for the objective function [Eq.~\eqref{eq:C_parity_mean}] yields a lower approximation ratio than using the best spanning tree energy $\tilde{\calC}^{\mathrm{best}}$ [Eq.~\eqref{eq:C_parity_min}].
Note that, when the mean spanning tree energy is used as the objective function, we still use the best spanning tree energy to obtain the best result after the optimization, in accordance with the methods in Ref.~\cite{Weidinger2023}. Adding more layers increases the performance. This is in contrast to Ref.~\cite{Wybo_Leib_Quantum2024} where trivial instances (i.e., instances that are solvable by classical state decoding) make up a major part of the simulated instances.

\begin{figure}
    \centering 
    \includegraphics[width=\columnwidth]{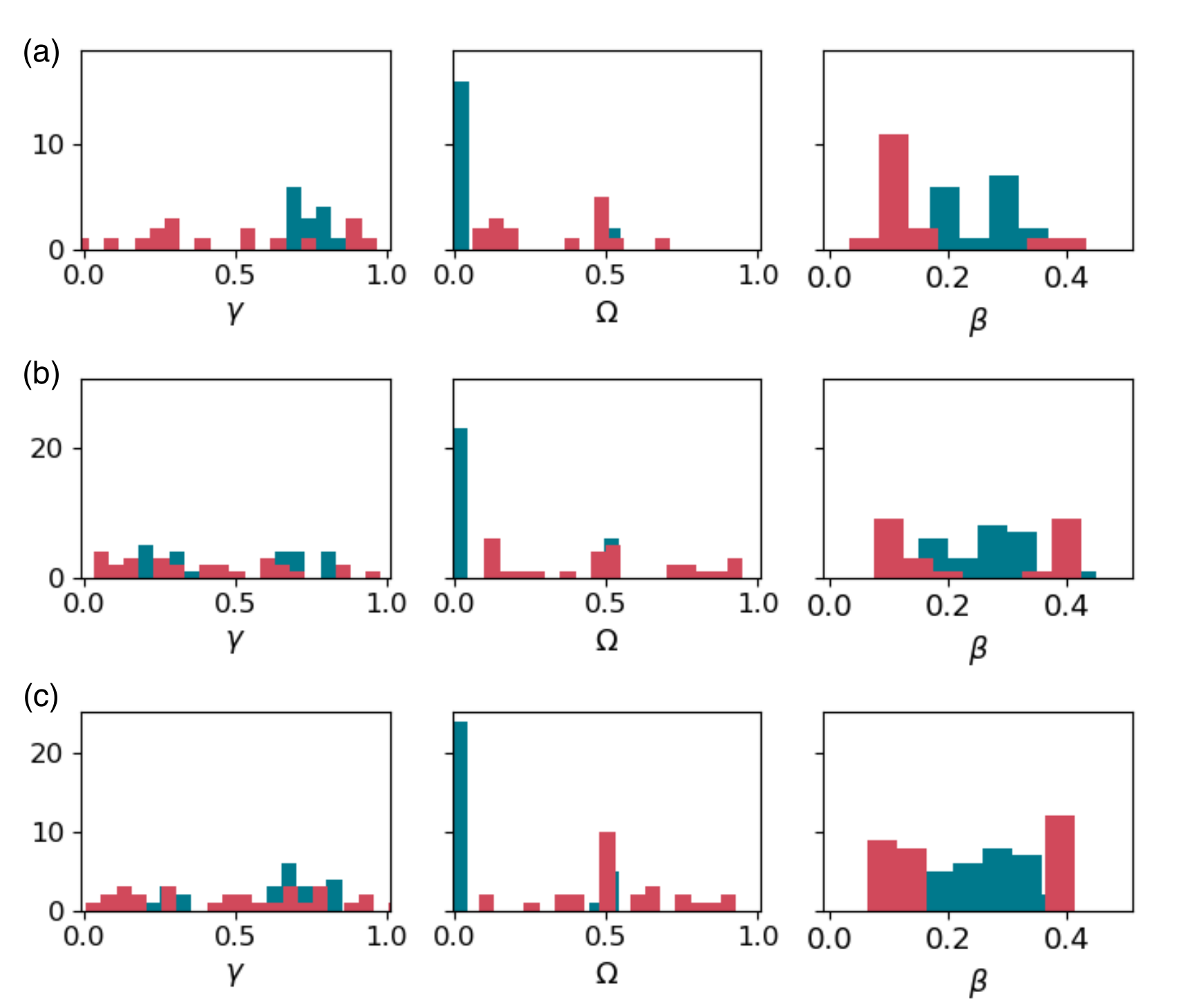}
    \caption{Histogram of the best parameters (in units of $\pi$) found in {\parity} QAOA using $\tilde{\calC}^{\mathrm{best}}$, ${p=2}$ and ${N=7}$ corresponding to the 3 different problem cases in Fig.~\ref{fig:app:Optimization_NotSolvedJij}: (a) non-trivial instances for ${p=5}$, (b) ${p=4}$ and (c) ${p=3}$. Blue indicates the parameters from the first ${(\gamma_1^*, \Omega_1^*, \beta_1^*)}$, and red indicates the parameters for the second layer, ${(\gamma_2^*, \Omega_2^*, \beta_2^*)}$. 
   }
    \label{fig:app:Parameter_NotSolvedJij}
\end{figure}

Figure~\ref{fig:app:Parameter_NotSolvedJij} shows the best parameters found by the classical optimization routine for Parity QAOA, with the$\tilde{\calC}^{\mathrm{best}}$ objective function, for ${p=2}$. Blue bars indicate the values of the first layer, ${(\gamma_1^*, \Omega_1^*,\beta_1^*)}$. For the constraint parameter $\Omega_1^*$, the values concentrate at $\Omega_1^*=0$. This is in line with Ref.~\cite{Wybo_Leib_Quantum2024}. However, $\Omega_2^*$ is nonzero. This suggests that one can skip the constraint gates for the first layer. Further work could investigate if starting with the local-field ground state $\boldsymbol{s_3}$ instead of an equal superposition of all computational basis states leads to similar results.

\begin{table}[]
    \centering
    \begin{tabular}{c|c}
       $t_1$  & $(0, 2), (0, 1), (0, 6), (0, 7), (3, 7), (3, 5), (4, 5)$ \\
       $t_2$  & $(0, 2), (0, 7), (5, 7), (4, 5), (4, 6), (3, 4), (1, 4)$ \\
       $t_3$  & $(3, 6), (3, 7), (2, 7), (5, 7), (1, 5), (0, 1), (1, 4)$ \\
       $t_4$  & $(2, 7), (2, 6), (1, 2), (0, 6), (3, 6), (1, 5), (4, 6)$ \\
       $t_5$  & $(2, 6), (1, 2), (3, 4), (3, 5), (0, 6), (3, 6), (2, 7)$ \\
       $t_6$  & $(0, 2), (0, 1), (2, 6), (1, 5), (5, 7), (3, 5), (4, 5)$ \\
       $t_7$  & $(1, 2), (0, 7), (1, 4), (1, 5), (3, 4), (3, 7), (4, 6)$ \\
       $t_8$  & $(1, 2), (0, 6), (0, 7), (0, 1), (3, 4), (3, 5), (3, 7)$
    \end{tabular}
    \caption{Spanning trees used for decoding for the 4-regular graph given in Fig.~\ref{fig:4-regular_graph}. With 8 logical qubits, 7 interactions (parity qubits) form a spanning tree. The tuple ${(i, j)}$ refers to the interaction between qubit $i$ and qubit $j$, i.e., to the parity qubit $ij$.}
    \label{tab:Spanning_Trees_4reg}
\end{table}

\begin{figure}
    \centering 
    \includegraphics[width=\columnwidth] {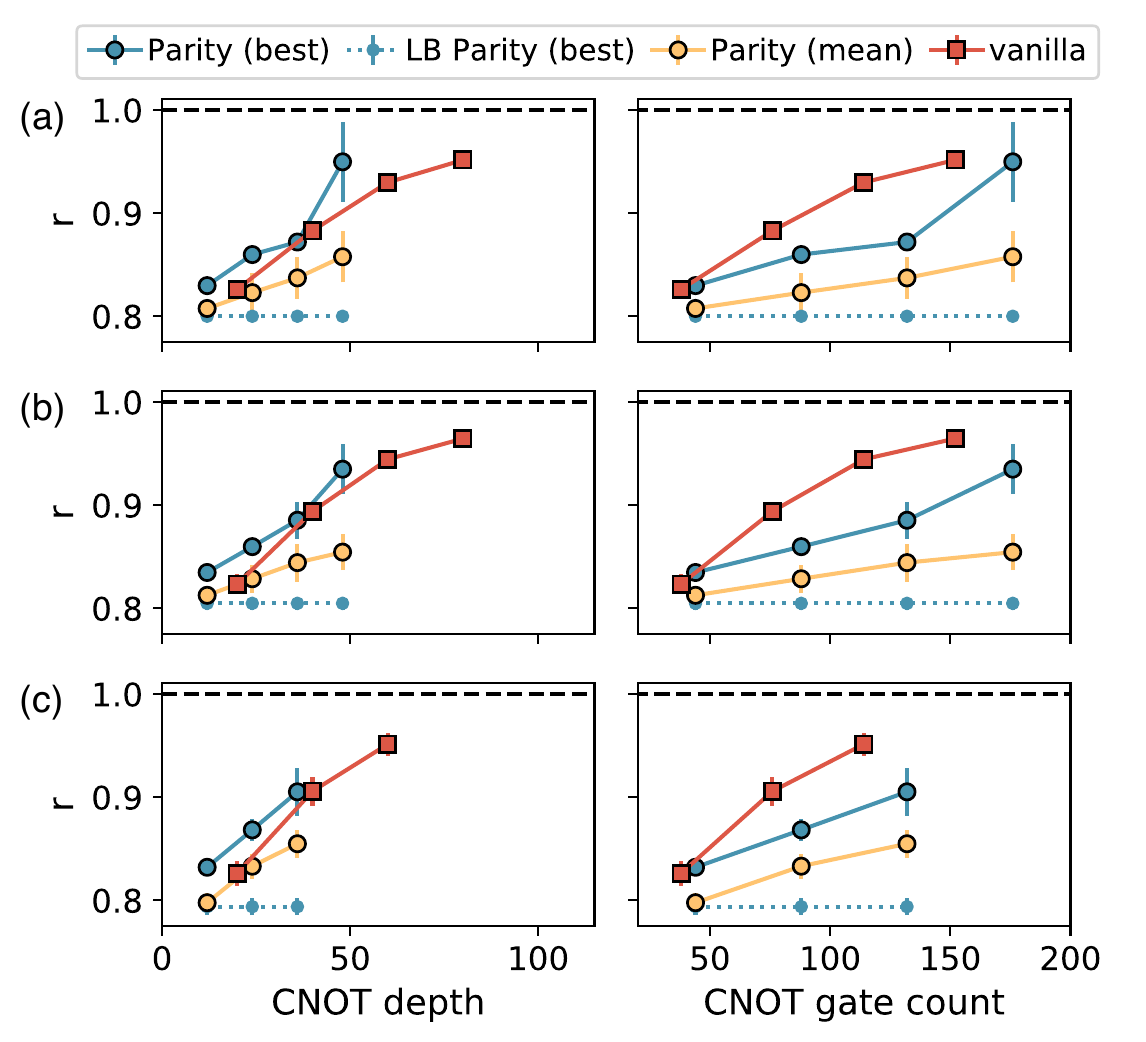}
    \caption{Approximation ratio $r$ of {\parity} and {\vanilla} QAOA for a 4-regular graph of all non-trivial instances for (a) ${p=5}$, (b) ${p=4}$ and (c) ${p=3}$ and $M=5$ in dependence of CNOT depth and CNOT gate count. In each line, the left-most data point represents ${p=1}$, the second left-most data point represents ${p=2}$, and so forth.
    The data points show the average of 6/14/16 non-trivial instances at ${p=5}$/${p=4}$/${p=3}$. That are all non-trivial instances for each case whereas no instance is shown double, e.g. instances included in (a) are not considered in (b) again.
    The error bars indicate two times the standard error of the mean and are not visible due to the size of the markers. Classical optimization parameters are: ${n_\text{init}=80}$, ${n_\text{mc}=400/1000/1500/2000}$ for ${p=1/2/3/4}$.
   }
    \label{fig:app:Optimization_NotSolvedJij_4reg}
\end{figure}
\subsection{4-regular graph}\label{sec:app:4regular_optimization}
Similar to the complete graph, we generate all unique problem instances for the 4-regular graph, shown in Fig.~\ref{fig:4-regular_graph}. This graph has $10$ constraints in the parity mapping, leading to ${2^{10}=1024}$ unique problem instances. The spanning trees for decoding are given in Tab.~\ref{tab:Spanning_Trees_4reg}. Note that the ancilla parity qubit $16$ is never included. Further, we consider two different sets, one with 5 trees ${T=\{t_1, t_2, t_3, ..., t_5\}}$ and one with 8 trees ${T=\{t_1, t_2, t_3, ..., t_8\}}$.
The explicit numbers of non-trivial instances for 5 spanning trees are given in Tab.~\ref{tab:app:Number_of_nontrivial_instances}. Numbers for 8 spanning trees are given in brackets. 
Figure~\ref{fig:app:SuccessRate_allJij}(b) shows the LB of the success rate of {\parity} QAOA for ${p\in\{1, 2, \dots 6\}}$ for both ${M=5}$ and ${M=8}$. The success rate is higher for the case with ${M=8}$, showing that reading out more spanning trees increases the chance of decoding to the ground state. Both lines converge to $1$, where the line for ${M=8}$ reaches this limit at ${p=5}$ while the line for ${M=5}$ reaches it at ${p=6}$. 
Due to the higher amount of non-trivial instances, we perform QAOA simulations for the non-trivial cases with ${M=5}$ spanning trees. The results are shown in Fig.~\ref{fig:app:Optimization_NotSolvedJij_4reg}. As before, we distinguish between non-trivial instances at ${p=5}$ (a), at ${p=4}$ (b) and at ${p=3}$ (c). Instances that are shown in one panel are not considered again in another panels. 
In line with previous results, {\parity} QAOA with $\tilde{\calC}^{\mathrm{best}}$ as objective function requires lower CNOT depth but a higher CNOT gate count. The LB of the approximation ratio for {\parity} QAOA does not increase with increasing layers but the performance of {\parity} QAOA does. Again, using the mean spanning tree energy $\tilde{\calC}^{\mathrm{mean}}$ [Eq.~\eqref{eq:C_parity_mean}] as the objective function does not perform as well as the best spanning tree energy approach $\tilde{\calC}^{\mathrm{best}}$ [Eq.~\eqref{eq:C_parity_min}].

\section{QAOA for Hypergraphs}\label{sec:app:hypergraph}

\begin{figure}
\includegraphics[width=\columnwidth]{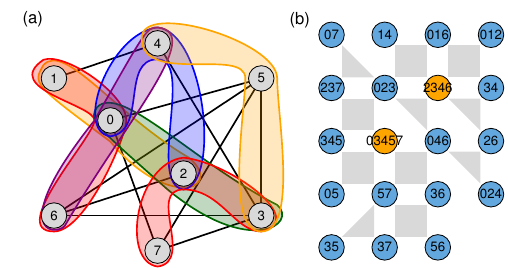}
    \caption{(a) Hypergraph in the logical representation. Hyperedges that connect 3 nodes are depicted as closed curves around the nodes and hyperedges that connect 2 nodes are depicted as lines between the nodes. (b) Compiled parity graph obtained from the hypergraph shown in (a) with two ancilla qubits (colored yellow). }
    \label{fig:app:hypergraph_visualization}
\end{figure}

{\Parity} QAOA can be applied to QUBO problems (as described in the main text, Sec.~\ref{sec:ParityQAOA}) and to higher-order binary optimization problems~\cite{Ender2023}. This section will focus on QAOA simulations for a problem with higher-order interaction terms, which is represented by the hypergraph shown in Fig.~\ref{fig:app:hypergraph_visualization}(a). The hypergraph consists of hyperedges that connect up to 3 nodes. Therefore, we modify Eq.~\eqref{eq:Hp-qubo} to take 3-body interaction into account and rewrite it as
\begin{equation}\label{eq:app:hyper_hamiltonian}
    H_P = \sum_{(i, j)\in E_2} J_{ij} \sigma_i^z \sigma_j^z + \sum_{(i, j, k)\in E_3} J_{ijk} \sigma_i^z \sigma_j^z \sigma_k^z,
\end{equation}
where $E_2$ is the set of edges that connect 2 nodes, $E_3$ is the set of hyperedges that connect 3 nodes, and $J_{ij}$ and $J_{ijk}$ denote the weights for the corresponding hyperedges for the given $i,j (,k)$ nodes. In Fig.~\ref{fig:app:hypergraph_visualization}(a) hyperedges that connect 2 nodes are depicted as lines (10 in total), and hyperedges that connect 3 nodes are depicted as closed curves (7 in total). In Fig.~\ref{fig:app:hypergraph_visualization}(b), one can find the corresponding parity mapping that was obtained by parity compilation~\cite{Ender2023, parityos}.

\begin{figure}
    \includegraphics[width=\columnwidth]{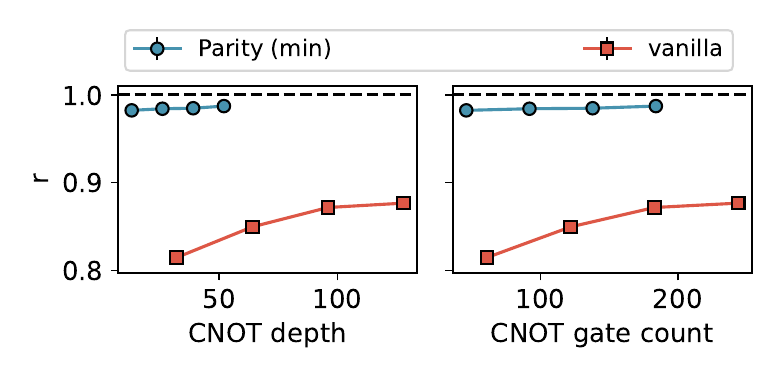}
    \caption{Approximation ratio $r$ for the given hypergraph for {\vanilla} and {\parity} QAOA as a function of the CNOT depth and CNOT gate count. In each line, the left-most data point represents ${p=1}$, the second left-most data point represents ${p=2}$, and so forth. 
    Results show the average over 100 simulated instances. As a classical optimizer, we used COBYLA from SciPy~\cite{2020SciPy-NMeth, powell1994direct, powell1998direct, powell2007view}, with 200 optimization steps, tolerance was set to $10^{-4}$, and ${n_\text{init}=25/50/75/100}$ for ${p=1/2/3/4}$. The number of circuit runs for the objective function evaluation was set to 500.}
    \label{fig:app:hypergraph_r_cnot}
\end{figure}

In Fig.~\ref{fig:app:hypergraph_r_cnot} we show the simulation results in terms of the approximation ratio for {\vanilla} and {\parity} QAOA for this hypergraph. Here, {\parity} QAOA is simulated with the best spanning tree energy $\tilde{\calC}^{\mathrm{best}}$, Eq.~\eqref{eq:C_parity_min}, as an objective function, using ${M = 5}$ readout bases obtained by the ParityOS package~\cite{parityos}. In this context, a readout basis is defined as a minimal set of physical qubits from which one can decode the configuration of the logical qubits and can be seen as a generalization of a spanning tree for hypergraphs.
For the simulations, 100 problems were randomly generated by assigning random weights from $\{{-1, +1\}}$ to the coefficients $J_{ij}$ and $J_{ijk}$ for each hyperedge in Eq.~\eqref{eq:app:hyper_hamiltonian}. The QAOA simulations presented in this section were performed with qiskit~\cite{Qiskit}. For hypergraphs, we used the COBYLA optimizer from SciPy~\cite{2020SciPy-NMeth, powell1994direct, powell1998direct, powell2007view} instead of the Monte-Carlo optimizer employed in the main text, because we found it to be more efficient on this problem class.

The results show that for this hypergraph, {\parity} QAOA outperforms {\vanilla} QAOA. Also, in contrast to the 4-regular graph case (see Fig.~\ref{fig:N8_optimization_4reg}), {\parity} QAOA requires fewer resources in terms of both, CNOT count and CNOT depth. To be precise, {\parity} QAOA has a CNOT depth (gate count) of 13 (46) and {\vanilla} QAOA has a CNOT depth (gate count) of 32 (61). The circuit for {\vanilla} QAOA was optimized with the qiskit~\cite{Qiskit} and tket compilers~\cite{Sivarajah_IOPPublishing2021}, where the qiskit compiler yielded better results, which we used for the simulations. The lower CNOT depth and lower gate count for {\parity} QAOA for this problem graph are in line with the results in Ref.~\cite{Fellner_Quantum2023}, an advantage of {\parity} QAOA in those metrics for hypergraphs is anticipated.

%

\end{document}